\pdfoutput=1

\documentclass[12pt,a4paper]{article}

\usepackage{ifthen} 
\newboolean{pdflatex}
\setboolean{pdflatex}{true} 

\newboolean{articletitles}
\setboolean{articletitles}{true} 

\newboolean{uprightparticles}
\setboolean{uprightparticles}{true} 

\newboolean{inbibliography}
\setboolean{inbibliography}{false} 

\usepackage{graphpap}
\usepackage{color}
\usepackage{rotating}
\usepackage{multirow}
\usepackage{verbatim}

\usepackage{microtype}
\usepackage{lineno}  
\usepackage{xspace} 
\usepackage{caption} 

\usepackage{graphicx}  
\usepackage{color}
\usepackage{colortbl}
\graphicspath{{./figs/}} 

\usepackage{amsmath} 
\usepackage{amssymb}
\usepackage{amsfonts}
\usepackage{upgreek} 

\newcommand*\patchAmsMathEnvironmentForLineno[1]{%
\expandafter\let\csname old#1\expandafter\endcsname\csname #1\endcsname
\expandafter\let\csname oldend#1\expandafter\endcsname\csname
end#1\endcsname
 \renewenvironment{#1}%
   {\linenomath\csname old#1\endcsname}%
   {\csname oldend#1\endcsname\endlinenomath}%
}
\newcommand*\patchBothAmsMathEnvironmentsForLineno[1]{%
  \patchAmsMathEnvironmentForLineno{#1}%
  \patchAmsMathEnvironmentForLineno{#1*}%
}
\AtBeginDocument{%
\patchBothAmsMathEnvironmentsForLineno{equation}%
\patchBothAmsMathEnvironmentsForLineno{align}%
\patchBothAmsMathEnvironmentsForLineno{flalign}%
\patchBothAmsMathEnvironmentsForLineno{alignat}%
\patchBothAmsMathEnvironmentsForLineno{gather}%
\patchBothAmsMathEnvironmentsForLineno{multline}%
\patchBothAmsMathEnvironmentsForLineno{eqnarray}%
}

\usepackage{hyperref}    
\usepackage[all]{hypcap} 

\def\lhcb {\mbox{LHCb}\xspace}
\def\atlas  {\mbox{ATLAS}\xspace}
\def\cms    {\mbox{CMS}\xspace}

\def\cdf    {\mbox{CDF}\xspace}

\def\rich   {RICH\xspace}

\def\spd    {SPD\xspace}
\def\presh  {PS\xspace}
\def\ecal   {ECAL\xspace}

\ifthenelse{\boolean{uprightparticles}}%
{
 
 \def\Pgamma      {\ensuremath{\upgamma}\xspace}

 \def\Pmu         {\ensuremath{\upmu}\xspace}

 \def\Ppi         {\ensuremath{\uppi}\xspace}

 \def\Pchi        {\ensuremath{\upchi}\xspace}                 
 \def\Ppsi        {\ensuremath{\uppsi}\xspace}

 \def\PDelta      {\ensuremath{\Delta}\xspace}                 
 \def\PXi      {\ensuremath{\Xi}\xspace}                 
 \def\PLambda      {\ensuremath{\Lambda}\xspace}                 
 \def\PSigma      {\ensuremath{\Sigma}\xspace}                 
 \def\POmega      {\ensuremath{\Omega}\xspace}                 
 \def\PUpsilon      {\ensuremath{\Upsilon}\xspace}                 
 
 \def\PB      {\ensuremath{\mathrm{B}}\xspace}                 
                  
 \def\PD      {\ensuremath{\mathrm{D}}\xspace}

 \def\PJ      {\ensuremath{\mathrm{J}}\xspace}                 
 \def\PK      {\ensuremath{\mathrm{K}}\xspace}

 \def\Pb      {\ensuremath{\mathrm{b}}\xspace}                 
 \def\Pc      {\ensuremath{\mathrm{c}}\xspace}

 \def\Pi      {\ensuremath{\mathrm{i}}\xspace}

 \def\Pp      {\ensuremath{\mathrm{p}}\xspace}

}
{
 
 \def\Pgamma      {\ensuremath{\gamma}\xspace}

 \def\Pmu         {\ensuremath{\mu}\xspace}

 \def\Ppi         {\ensuremath{\pi}\xspace}

 \def\Pchi        {\ensuremath{\chi}\xspace}                 
 \def\Ppsi        {\ensuremath{\psi}\xspace}                 
                  
 \mathchardef\PDelta="7101
 \mathchardef\PXi="7104
 \mathchardef\PLambda="7103
 \mathchardef\PSigma="7106
 \mathchardef\POmega="710A
 \mathchardef\PUpsilon="7107
                  
 \def\PB      {\ensuremath{B}\xspace}                 
                  
 \def\PD      {\ensuremath{D}\xspace}

 \def\PJ      {\ensuremath{J}\xspace}                 
 \def\PK      {\ensuremath{K}\xspace}

 \def\Pb      {\ensuremath{b}\xspace}                 
 \def\Pc      {\ensuremath{c}\xspace}

 \def\Pi      {\ensuremath{i}\xspace}

 \def\Pp      {\ensuremath{p}\xspace}

}

\makeatletter
\ifcase \@ptsize \relax
  \newcommand{\miniscule}{\@setfontsize\miniscule{4}{5}}
\or
  \newcommand{\miniscule}{\@setfontsize\miniscule{5}{6}}
\or
  \newcommand{\miniscule}{\@setfontsize\miniscule{5}{6}}
\fi
\makeatother

\DeclareRobustCommand{\optbar}[1]{\shortstack{{\miniscule (\rule[.5ex]{1.25em}{.18mm})}
  \\ [-.7ex] $#1$}}

\def\mumu       {{\ensuremath{\Pmu^+\Pmu^-}}\xspace}

\def\g      {{\ensuremath{\Pgamma}}\xspace}

\def\cquark    {{\ensuremath{\Pc}}\xspace}

\def\bquark    {{\ensuremath{\Pb}}\xspace}

\def\pion   {{\ensuremath{\Ppi}}\xspace}
\def\piz    {{\ensuremath{\pion^0}}\xspace}

  \def\Kbar    {{\kern 0.2em\overline{\kern -0.2em \PK}{}}\xspace}

\def\KorKbar    {\kern 0.18em\optbar{\kern -0.18em K}{}\xspace}

  \def\Dbar    {{\kern 0.2em\overline{\kern -0.2em \PD}{}}\xspace}

\def\DorDbar    {\kern 0.18em\optbar{\kern -0.18em D}{}\xspace}

\def\B       {{\ensuremath{\PB}}\xspace}
\def\Bbar    {{\ensuremath{\kern 0.18em\overline{\kern -0.18em \PB}{}}}\xspace}

\def\BorBbar    {\kern 0.18em\optbar{\kern -0.18em B}{}\xspace}

\def\Bu      {{\ensuremath{\B^+}}\xspace}

\def\jpsi     {{\ensuremath{{\PJ\mskip -3mu/\mskip -2mu\Ppsi\mskip 2mu}}}\xspace}

\def\Y#1S{\ensuremath{\PUpsilon\mathrm{(#1S)}}\xspace}

\def\chic  {{\ensuremath{\Pchi_{\cquark}}}\xspace}

\def\proton      {{\ensuremath{\Pp}}\xspace}

\def\Lbar        {{\ensuremath{\kern 0.1em\overline{\kern -0.1em\PLambda}}}\xspace}
\def\LorLbar    {\kern 0.18em\optbar{\kern -0.18em \PLambda}{}\xspace}

\def\BF         {{\ensuremath{\cal B}}\xspace}

\def\BR         {\BF}
\newcommand{\decay}[2]{\ensuremath{#1\!\to #2}\xspace}         

\def\to                 {\ensuremath{\rightarrow}\xspace}

\def\AT#1     {\ensuremath{A_{\mathrm{T}}^{#1}}\xspace}           

\def\C#1      {\ensuremath{\mathcal{C}_{#1}}\xspace}                       
\def\Cp#1     {\ensuremath{\mathcal{C}_{#1}^{'}}\xspace}                    
\def\Ceff#1   {\ensuremath{\mathcal{C}_{#1}^{\mathrm{(eff)}}}\xspace}        
\def\Cpeff#1  {\ensuremath{\mathcal{C}_{#1}^{'\mathrm{(eff)}}}\xspace}       
\def\Ope#1    {\ensuremath{\mathcal{O}_{#1}}\xspace}                       
\def\Opep#1   {\ensuremath{\mathcal{O}_{#1}^{'}}\xspace}                    

\newcommand{\tev}{\ifthenelse{\boolean{inbibliography}}{\ensuremath{~T\kern -0.05em eV}\xspace}{\ensuremath{\mathrm{\,Te\kern -0.1em V}}}\xspace}
\newcommand{\gev}{\ensuremath{\mathrm{\,Ge\kern -0.1em V}}\xspace}
\newcommand{\mev}{\ensuremath{\mathrm{\,Me\kern -0.1em V}}\xspace}
\newcommand{\kev}{\ensuremath{\mathrm{\,ke\kern -0.1em V}}\xspace}
\newcommand{\ev}{\ensuremath{\mathrm{\,e\kern -0.1em V}}\xspace}
\newcommand{\gevc}{\ensuremath{{\mathrm{\,Ge\kern -0.1em V\!/}c}}\xspace}
\newcommand{\mevc}{\ensuremath{{\mathrm{\,Me\kern -0.1em V\!/}c}}\xspace}
\newcommand{\gevcc}{\ensuremath{{\mathrm{\,Ge\kern -0.1em V\!/}c^2}}\xspace}
\newcommand{\gevgevcccc}{\ensuremath{{\mathrm{\,Ge\kern -0.1em V^2\!/}c^4}}\xspace}
\newcommand{\mevcc}{\ensuremath{{\mathrm{\,Me\kern -0.1em V\!/}c^2}}\xspace}

\def\mum  {\ensuremath{{\,\upmu\rm m}}\xspace}

\def\invfb   {\ensuremath{\mbox{\,fb}^{-1}}\xspace}

\newcommand{\chisq}{\ensuremath{\chi^2}\xspace}

\def\gsim{{~\raise.15em\hbox{$>$}\kern-.85em
          \lower.35em\hbox{$\sim$}~}\xspace}
\def\lsim{{~\raise.15em\hbox{$<$}\kern-.85em
          \lower.35em\hbox{$\sim$}~}\xspace}

\def\sqs   {\ensuremath{\protect\sqrt{s}}\xspace}

\def\pt         {\mbox{$p_{\rm T}$}\xspace}

\def\evtgen     {\mbox{\textsc{EvtGen}}\xspace}

\def\geant      {\mbox{\textsc{Geant4}}\xspace}

\def\photos     {\mbox{\textsc{Photos}}\xspace}

\def\pythia     {\mbox{\textsc{Pythia}}\xspace}

\def\tell1  {TELL1\xspace}
\def\ukl1   {UKL1\xspace}

\newcommand{\eg}{\mbox{\itshape e.g.}\xspace}

\def\chib           {\ensuremath{\Pchi_{\bquark}}\xspace}

\def\chibone        {\ensuremath{\Pchi_{\bquark 1}}\xspace}
\def\chibtwo        {\ensuremath{\Pchi_{\bquark 2}}\xspace}

\def\chibOneP       {\ensuremath{\Pchi_{\bquark}\mathrm{(1P)}}\xspace}
\def\chibTwoP       {\ensuremath{\Pchi_{\bquark}\mathrm{(2P)}}\xspace}
\def\chibThreeP     {\ensuremath{\Pchi_{\bquark}\mathrm{(3P)}}\xspace}

\def\chiboneOneP    {\ensuremath{\Pchi_{\bquark 1}\mathrm{(1P)}}\xspace}
\def\chiboneTwoP    {\ensuremath{\Pchi_{\bquark 1}\mathrm{(2P)}}\xspace}
\def\chiboneThreeP  {\ensuremath{\Pchi_{\bquark 1}\mathrm{(3P)}}\xspace}

\def\chibtwoOneP    {\ensuremath{\Pchi_{\bquark 2}\mathrm{(1P)}}\xspace}
\def\chibtwoTwoP    {\ensuremath{\Pchi_{\bquark 2}\mathrm{(2P)}}\xspace}
\def\chibtwoThreeP  {\ensuremath{\Pchi_{\bquark 2}\mathrm{(3P)}}\xspace}

\def\ups            {\ensuremath{\PUpsilon}\xspace}
\def\YnS            {\ensuremath{\PUpsilon\mathrm{(nS)}}\xspace}

\def\chibonep       {\ensuremath{\Pchi_{\bquark}\mathrm{(1P)}}\xspace}
\def\chibtwop       {\ensuremath{\Pchi_{\bquark}\mathrm{(2P)}}\xspace}
\def\chibthreep     {\ensuremath{\Pchi_{\bquark}\mathrm{(3P)}}\xspace}
\def\chibmp         {\ensuremath{\Pchi_{\bquark}\mathrm{(mP)}}\xspace}

\def\Rmn           {\ensuremath{\mathcal{R}^{\chibmp}_{\YnS}}\xspace}

\usepackage{cite} 
\usepackage{mciteplus}

\usepackage{booktabs}
\usepackage{subfigure}
\usepackage{longtable} 
\usepackage{pifont}
\usepackage{amssymb}

\begin{document}

\renewcommand{\thefootnote}{\fnsymbol{footnote}}
\setcounter{footnote}{1}


\begin{titlepage}
\pagenumbering{roman}

\vspace*{-1.5cm}
\centerline{\large EUROPEAN ORGANIZATION FOR NUCLEAR RESEARCH (CERN)}
\vspace*{1.5cm}
\hspace*{-0.5cm}
\begin{tabular*}{\linewidth}{lc@{\extracolsep{\fill}}r}
\ifthenelse{\boolean{pdflatex}}
{\vspace*{-2.7cm}\mbox{\!\!\!\includegraphics[width=.14\textwidth]{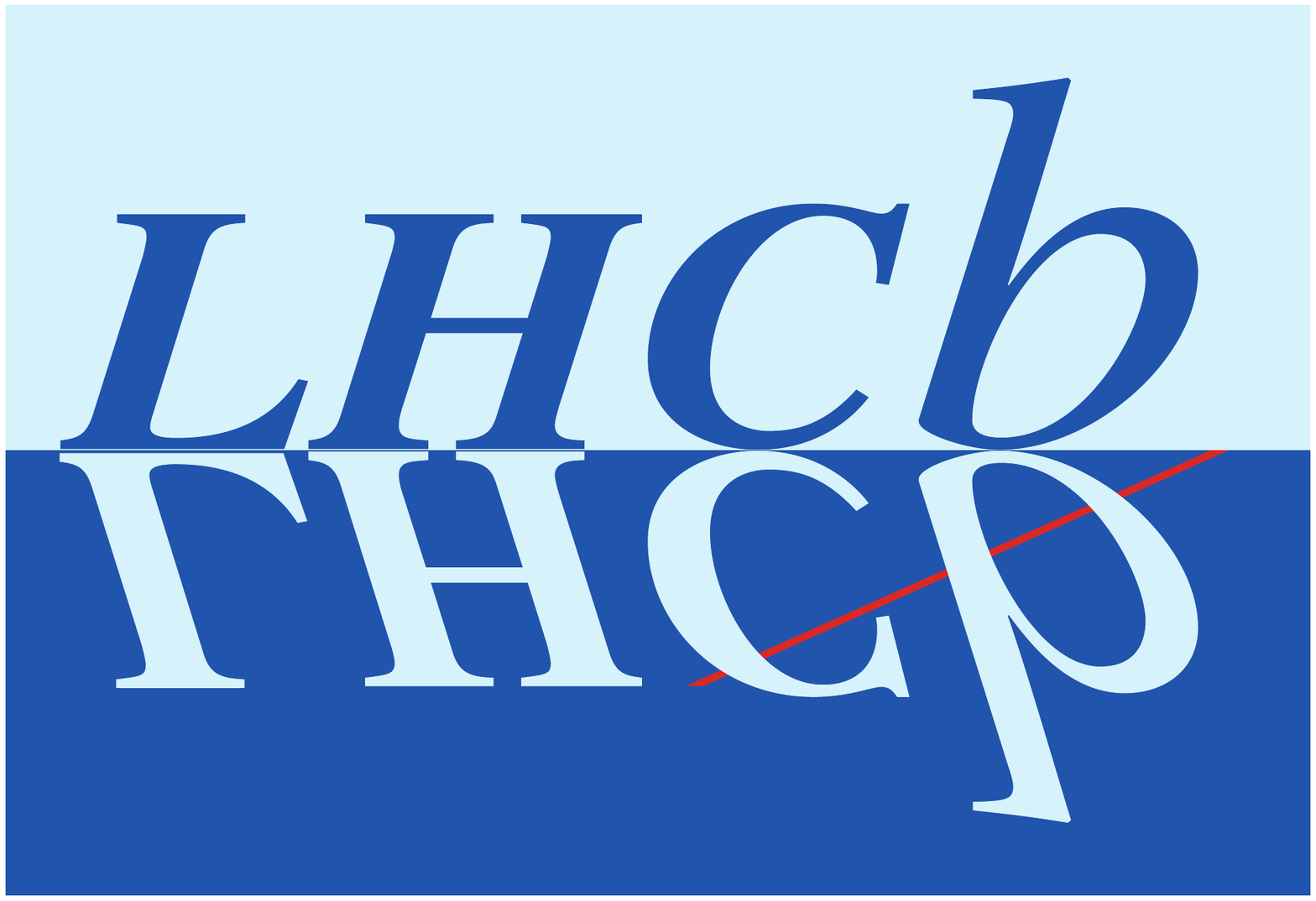}} & &}%
{\vspace*{-1.2cm}\mbox{\!\!\!\includegraphics[width=.12\textwidth]{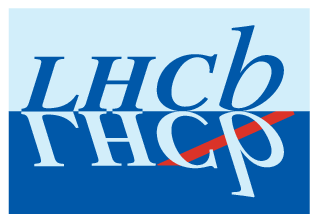}} & &}%
\\
   & & CERN-PH-EP-2014-178 \\  
   & & LHCb-PAPER-2014-031 \\  
   & & July 29, 2014  
\end{tabular*}

\vspace*{1.5cm}

{\bf\boldmath\huge
\begin{center}
  Study of \chib~meson
  production
  in \proton\proton~collisions at \mbox{$\sqrt{s}=7$}~and~\mbox{$8\tev$}
  and observation of the~decay \mbox{$\chibThreeP\to\Y3S\g$}
\end{center}
}

\vspace*{1.5cm}

\begin{center}
The LHCb collaboration\footnote{Authors are listed at the end of this paper.}
\end{center}

\vspace{\fill}

\begin{abstract}
  \noindent
  A study of \chib~meson production at \lhcb is performed on proton-proton collision data,
  corresponding to~3.0\invfb  of integrated luminosity 
  collected at centre-of-mass energies  \sqs = 7 and 8 \tev.
  The fraction of \YnS~mesons originating from \chib~decays is measured as a function of
  the~\ups~transverse momentum
  in the~rapidity range $2.0 < y^{\PUpsilon} < 4.5$.
  The~radiative transition of the \chibThreeP~meson to \Y3S~is observed for the~first time.
  The~\chiboneThreeP mass is determined to be
  \begin{equation*}
      m_{\chiboneThreeP} = 10\,511.3 \pm 1.7 \pm 2.5\mevcc,
  \end{equation*}
  where the first uncertainty is statistical and the second is  systematic.
\end{abstract}

\vspace*{1.5cm}

\begin{center}
  Submitted to Eur.~Phys.~J.~C 
\end{center}

\vspace{\fill}

{\footnotesize 
\centerline{\copyright~CERN on behalf of the \lhcb collaboration, license \href{http://creativecommons.org/licenses/by/4.0/}{CC-BY-4.0}.}}
\vspace*{2mm}

\end{titlepage}


\newpage
\setcounter{page}{2}
\mbox{~}



\renewcommand{\thefootnote}{\arabic{footnote}}
\setcounter{footnote}{0}



\pagestyle{plain} 
\setcounter{page}{1}
\pagenumbering{arabic}


%


\section{Introduction}
\label{sec:introduction}

The production of quarkonia states in high-energy hadron collisions
is described
in the~framework of 
non-relativistic quantum chromodynamics\,(NRQCD),
as two-step process:
a~heavy quark-antiquark pair is first created perturbatively at short distances, then it
evolves non-perturbatively into quarkonium at long distances.
The~NRQCD framework makes use of a combination of colour-singlet  and colour-octet
mechanisms~\cite{Kartvelishvili:1978id,Berger:1980ni,Baier:1981uk,Braaten:1994vv,Bodwin:1994jh}.
Recent calculations~\cite{Gong:2008sn,Campbell:2007ws,Artoisenet:2008fc,Lansberg:2008gk,Ma:2010vd}
support the~leading role of the~colour-singlet mechanism.
The comparison of experimental data for prompt production of \mbox{S-wave} quarkonia,
{\it{e.g.}}  \jpsi~or~\Y1S~mesons, with theory predictions requires
knowledge of feed-down contributions from P-wave quarkonia
states, \eg~radiative \decay{\chib}{\ups\g}~decays.
This contribution could significantly influence
the~interpretation of the~measured polarization of S-wave vector
quarkonia. In addition, measurements of the~relative production rates of P-wave to S-wave
quarkonia, as well as the~tensor-to-vector ratios, provide valuable
information on colour-octet matrix elements~\cite{Ma:2010vd,Likhoded:2012hw,Likhoded:2013aya}.

The production of P-wave charmonia,
jointly refered to as \chic~states,
has been studied  by
the~\cdf~\cite{Abulencia:2007bra}, \mbox{HERA-B}~\cite{Abt:2008ed}
and \lhcb~\cite{LHCb-PAPER-2011-019,LHCb-PAPER-2011-030,LHCb-PAPER-2013-028} collaborations;
measurements involving \chib~states
have been performed by
the~\cdf~\cite{Affolder:1999wm},
\atlas~\cite{Aad:2011ih},
\cms~\cite{CMS-PAS-BPH-13-005} and
\lhcb~\cite{LHCb-PAPER-2012-015,LHCb-CONF-2012-020}
experiments.

This paper presents a measurement of the fractions of \ups~mesons
originating from radiative decays  of \chib mesons.
Depending on the~relative orientation of the~quark spins,
the~\chib~states can be either scalar, vector or tensor mesons, denoted by
$\Pchi_{\bquark \mathrm{J}}$ with total angular momentum $\mathrm{J}=0,1,2$.
The~analysis proceeds through the reconstruction of \ups~candidates via their dimuon decays.
The~fractions of \YnS~decays
originating from \chibmp~decays,
where $\mathrm{n}$~and~$\mathrm{m}$ are radial quantum numbers of
the~bound states are defined as
\begin{equation}
  \mathcal{R}^{\chibmp}_{\YnS} \equiv
  \dfrac
      {\upsigma  \left( \proton\proton \to \Pchi_{\bquark1}\mathrm{(mP)} \mathrm{X} \right)}
      {\upsigma  \left( \proton\proton \to \YnS    \mathrm{X} \right)}
      \times \BR_1
      + \dfrac
      {\upsigma  \left( \proton\proton \to \Pchi_{\bquark2}\mathrm{(mP)} \mathrm{X} \right)}
      {\upsigma  \left( \proton\proton \to \YnS    \mathrm{X} \right)}
      \times \BR_2
      \label{eq:r},
\end{equation}
where $\BR_{1(2)}$ denotes the branching fraction 
for the~decay \decay{\Pchi_{\bquark1(2)}\mathrm{(mP)}}{\YnS \g}.
Possible contributions from \decay{\Pchi_{\bquark0}\mathrm{(mP)}}{\YnS \g}~decays 
are neglected because  of the~small branching fraction for the~corresponding
radiative decays~\cite{PDG2012}.

The~results presented in this paper supersede earlier \lhcb~measurements~\cite{LHCb-PAPER-2012-015,LHCb-CONF-2012-020}.
In particular, the full data sample collected by \lhcb at \mbox{$\sqs=7$}~and 8\tev has been used and
the~measured fractions $\mathcal{R}^{\chibmp}_{\YnS}$
are reported for all six kinematically allowed transitions:
\mbox{\decay{\chibonep}{\Y1S\g}},
\mbox{\decay{\chibtwop}{\Y1S\g}},
\mbox{\decay{\chibtwop}{\Y2S\g}},
\mbox{\decay{\chibthreep}{\Y1S\g}},
\mbox{\decay{\chibthreep}{\Y2S\g}} and 
\mbox{\decay{\chibthreep}{\Y3S\g}}
in bins of transverse momentum of the~\ups~mesons in the rapidity range  $2.0<y<4.5$.
The last transition, which is usually not considered in theory predictions,
is observed for the first time.   
A~precise  measurement of the mass of the~\chiboneThreeP~meson,
which was recently
observed by
the~\atlas~\cite{Aad:2011ih}, D0~\cite{Abazov:2012gh} and
\lhcb~\cite{LHCb-CONF-2012-020} collaborations, is also performed. 

 \section{The \lhcb detector and data samples}
\label{sec:Detector}
The \lhcb detector~\cite{Alves:2008zz} is a single-arm forward
spectrometer covering the \mbox{pseudorapidity} range $2<\eta <5$,
designed for the study of heavy-flavoured particles. 
The detector includes a high-precision tracking system
consisting of a silicon-strip vertex detector surrounding the 
interaction region, a large-area silicon-strip detector located
upstream of a dipole magnet with a bending power of about
$4{\rm\,Tm}$, and three stations of silicon-strip detectors and straw
drift tubes placed downstream of the magnet.
The combined tracking system provides a momentum measurement with
a relative uncertainty that varies from 0.4\% at low momentum to 0.6\% at 100\gevc,
and an impact parameter measurement with a~resolution of 20\mum for
charged particles with large transverse momentum, \pt. Different types of charged hadrons are distinguished using information
from two ring-imaging Cherenkov detectors\,(\rich)~\cite{LHCb-DP-2012-003}.
Photon, electron and
hadron candidates are identified by a~calorimeter system consisting of
scintillating-pad\,(\spd) and preshower\,(\presh) detectors,
an~electromagnetic calorimeter\,(\ecal) and
a~hadronic calorimeter~\cite{LHCb-DP-2013-004}.
Muons are identified by a
system composed of alternating layers of iron and multiwire
proportional chambers~\cite{LHCb-DP-2012-002}.
The trigger~\cite{LHCb-DP-2012-004} consists of a
hardware stage, based on information from the calorimeter and muon
systems, followed by a software stage, which applies a full event
reconstruction.

Candidate events used in this analysis must pass the hardware trigger, with the specific requirement that the
product of the~\pt of two muon candidates be greater than
$(1.3\gevc)^2$ and $(1.6\gevc)^2$ for data collected
at~\mbox{$\sqs=7$}~and 8\tev,  respectively.
The~first stage of the~software trigger selects candidate events  
with two well-reconstructed tracks with hits
in the muon system,
\pt~greater than~500\mevc
and  momentum greater than~6\gevc
for each track.
The~two tracks are required 
to originate from
a~common vertex and
to~have an~invariant mass greater than~2.7\gevcc.
Events are required to pass a~second software trigger stage, where 
the previous trigger decision is confirmed using
improved track reconstruction algorithms,
and the requirement that the invariant mass of the dimuon pair exceeds 4.7\gevcc is 
applied.

The data samples used in this paper have been collected by
the~\lhcb~detector
in \proton\proton~collisions at~\mbox{$\sqs=7$}~and~8\tev 
with integrated luminosities of 1.0\invfb and 2.0\invfb, respectively. 
Simulated samples are used to determine signal efficiencies. In these 
samples, \PUpsilon~and \chib~mesons are produced unpolarized.
The effect of the unknown initial polarization on the~efficiencies,
and therefore on the results, is taken into account
as a~systematic uncertainty. 
In the simulation, \proton\proton~collisions are generated using
\pythia~\cite{Sjostrand:2006za}
with a specific \lhcb
configuration~\cite{LHCb-PROC-2010-056}.  Decays of hadrons
are described by \evtgen~\cite{Lange:2001uf}, in which final-state
radiation is generated using \photos~\cite{Golonka:2005pn}. The
interaction of the generated particles with the detector and its
response are implemented using the \geant
toolkit~\cite{Allison:2006ve, *Agostinelli:2002hh} as described in
Ref.~\cite{LHCb-PROC-2011-006}.
A~comparison of the distributions of the relevant variables used in this analysis is 
performed on data and simulated samples, in order to assess the reliability of
the~simulation 
in computing signal efficiencies and good agreement is found.

\section{Event selection and signal extraction}
\label{sec:Selection}

This~analysis proceeds through the~reconstruction of \YnS~candidates
via their dimuon decaysand their subsequent pairing with
a~photon candidate
to reconstruct
\mbox{\decay{\chib}{\PUpsilon\g}}~decays.

The~\ups~candidates are selected from pairs of oppositely charged tracks
identified as muons and originating from a~common vertex.
The~muons are required to have \pt larger than 1\gevc.
Good track quality is ensured by requiring a~$\chisq$
per degree of freedom, $\chisq/\rm{ndf}$,
of the~track fit to be less than~4~\cite{LHCb-DP-2013-003}.
A~multivariate estimator, based on information from the~tracking, muon and \rich~systems,
as well as compatibility with the~hypothesis of a~minimum ionizing particle in
the~calorimeter system~\cite{Powell,Terrier:691743,LHCb-DP-2013-001}, is used to improve 
the~muon identification purity.
The~identification efficiency for muons from $\ups\to\mumu$~decays
rises from 75\% to 98\%
as the~transverse momentum of the~muon increases from
1\gevc to 3\gevc.
A~good quality of the~two-prong common vertex is ensured by requiring
the~p-value of the~common
vertex fit to be greater than 0.5\%.
To~improve the~dimuon mass resolution and
to~suppress combinatorial
background from  muons originating in semileptonic decays of heavy-flavoured hadrons,
the~dimuon vertex is refitted using the~position of 
the~reconstructed \proton\proton~collision vertex as
an~additional constraint~\cite{Hulsbergen:2005pu}.
The~p-value for this fit is required to be larger
than 0.05\%.
When several collision vertices are reconstructed in the~event,
the~one closest to the~dimuon vertex is used.

\begin{figure*}[t!]
  \setlength{\unitlength}{1mm}
  \centering
  \begin{picture}(150,60)
    \put(0,0){
      \includegraphics*[width=75mm, height=60mm]{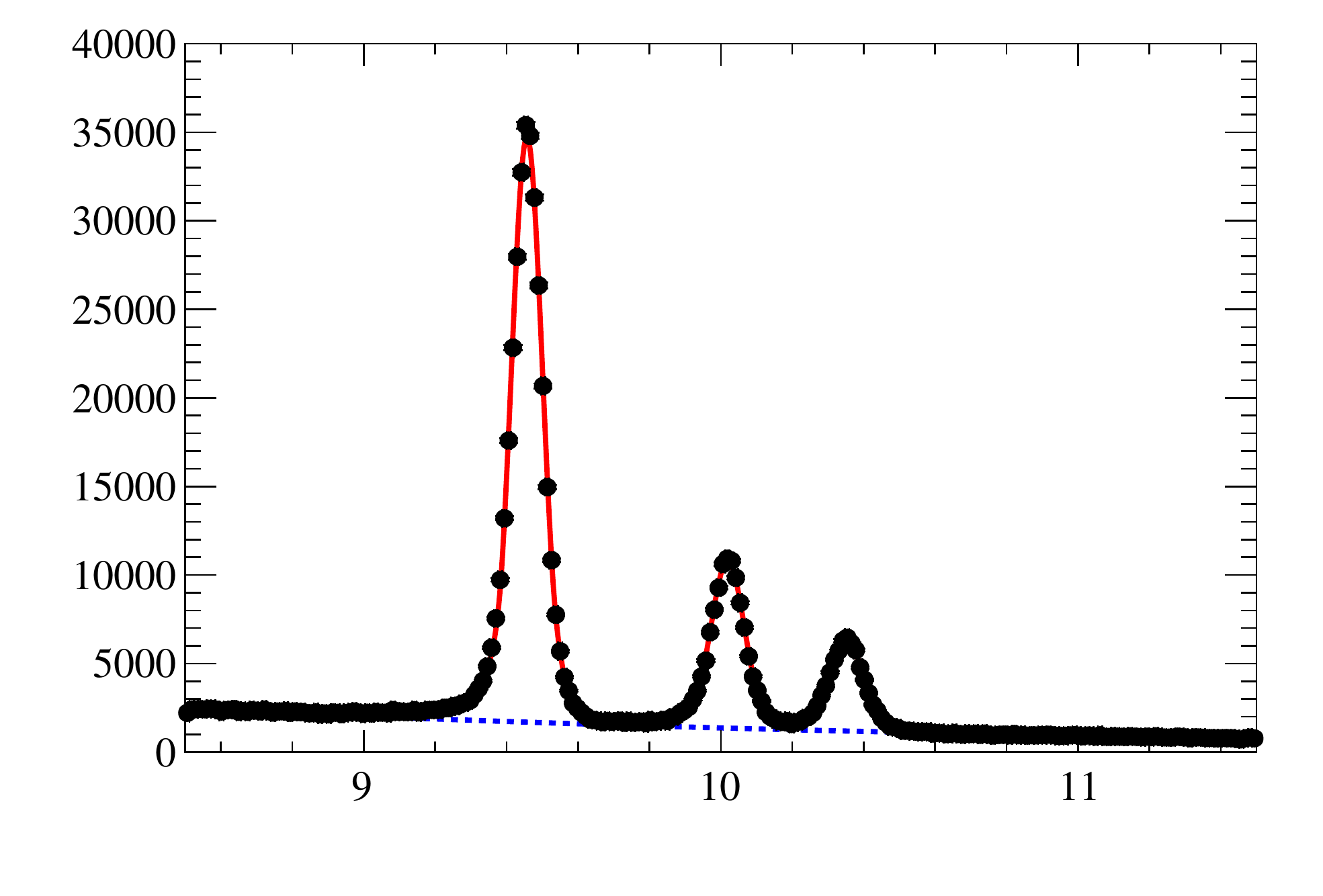}
    }
    \put(0,15){\small \begin{sideways}Candidates/(12\mevcc)\end{sideways}}
    \put(36, 2){$m_{\mumu}$}
    \put(58, 2){$\left[\gevcc\right]$}
    \put(45,47){$\begin{array}{l} \text{LHCb}\\ \sqs = 7 \tev \end{array}$}
    \put(75,0){
      \includegraphics*[width=75mm, height=60mm]{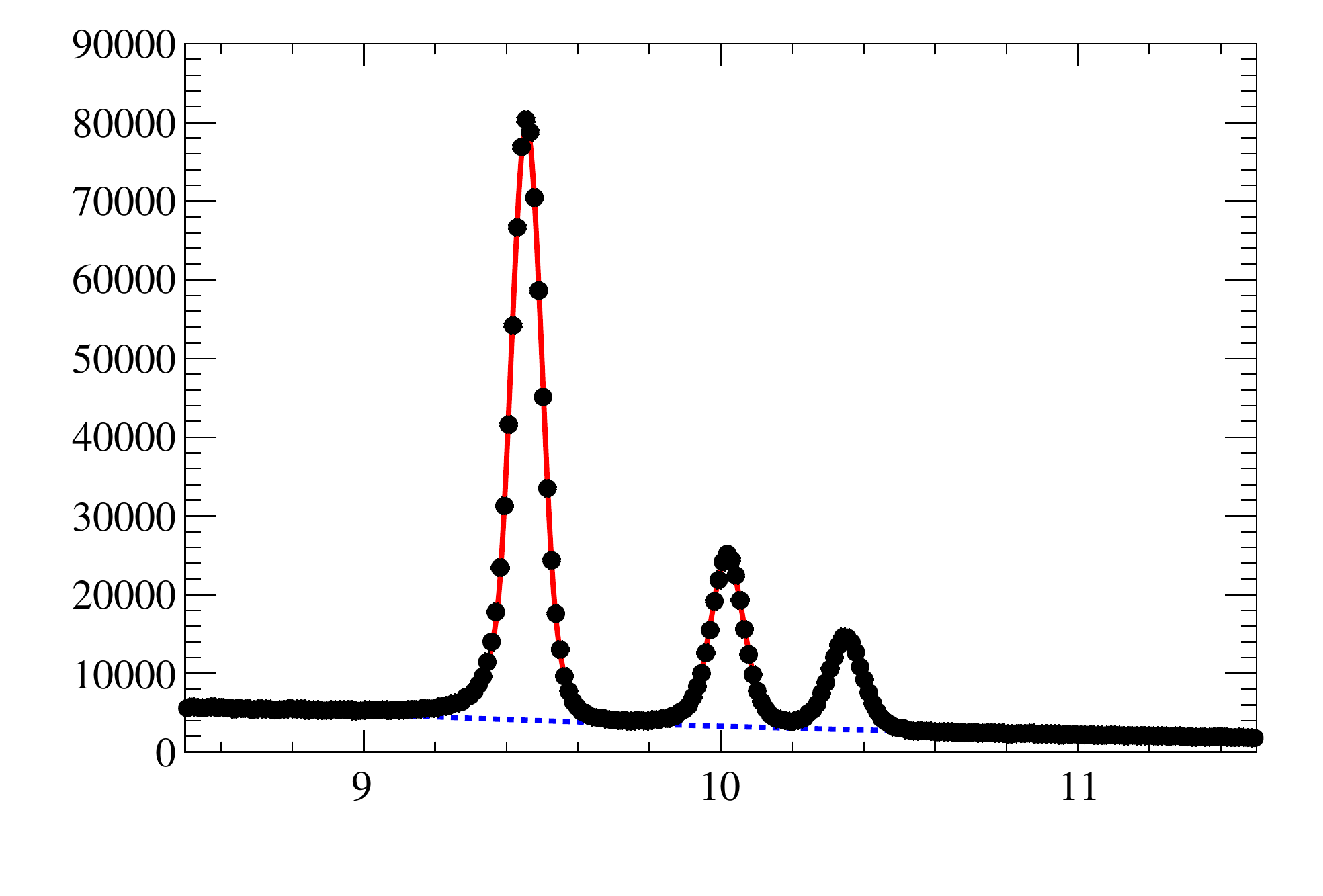}
    }
    \put(75,15){\small \begin{sideways}Candidates/(12\mevcc)\end{sideways}}
    \put(111,2){$m_{\mumu}$}
    \put(133,2){$\left[\gevcc\right]$}
    \put(120,47){$\begin{array}{l} \text{LHCb}\\ \sqs = 8 \tev \end{array}$}
    %
  \end{picture}
  \caption {\small
    Invariant mass distributions for selected dimuon candidates
    in the~kinematic range \mbox{$6<p_{\mathrm{T}}^{\mumu}<40\gevc$}
    and \mbox{$2.0<y^{\mumu}<4.5$} for (left)~data collected
    at~\mbox{$\sqs=7\,\mathrm{TeV}$}  and (right)~$8\,\mathrm{TeV}$.
    The three peaks on each plot
    correspond to the~\Y1S, \Y2S and \Y3S~signals\,(low-mass to high-mass).
    The~result of the~fit, described in the~text, is illustrated with
    a~red solid line, while the~background component is shown with
    a~blue dashed line.}
    \label{fig:upsilon:result:nominal}
\end{figure*}

The~invariant mass distributions for selected dimuon candidates
in the~kinematic range
of transverse momentum
\mbox{$6<p_{\mathrm{T}}^{\mumu}<40\gevc$}
and
rapidity 
\mbox{$2.0<y^{\mumu}<4.5$} are shown in
Fig.~\ref{fig:upsilon:result:nominal} for
data collected at $\sqs=7$~and 8\tev.
Three clear peaks are visible, corresponding to the~\Y1S, \Y2S and \Y3S signals\,(low-mass to high-mass).
The~yields of the~\YnS~signals are determined
using an~extended maximum  likelihood fit
to the~unbinned dimuon mass distributions.
The~fit function is parameterised as
the~sum of three signal
components and combinatorial background. 
Each \ups~signal has been modelled
with a modified Gaussian function with power-law tails on both
sides. 
The~combinatorial background is modelled with an~exponential function.
The~tail~parameters of the~signal functions are fixed using
simulated events, whereas the~mean and resolution
are allowed to vary in the~fit.
The~fit results are superimposed in Fig.~\ref{fig:upsilon:result:nominal}
and the~signal fit parameters are summarized in Table~\ref{tab:upsilon:result:nominal}.
The~peak~positions  and mass resolutions are found to be in
good agreement for the~data collected at~\mbox{$\sqs=7$}~and
8\tev, and in agreement with
the~known \YnS~masses~\cite{PDG2012}
and the~resolutions expected from simulated samples.

\begin{table}[t!]
  \centering 
  \caption{\small Yields of \YnS~mesons, 
    determined by fitting 
    the~dimuon invariant mass in
    the~range
    $6<p_{\mathrm{T}}^{\mumu}<40\gevc$ and
    $2.0<y^{\mumu}<4.5$, for  data  collected at~\mbox{$\sqs=7$}~and $8\,\mathrm{TeV}$.
    Only statistical uncertainties are shown.
  }\label{tab:upsilon:result:nominal}
  \vspace*{3mm}
  \begin{tabular*}{0.65\textwidth}{@{\hspace{10mm}}l@{\extracolsep{\fill}}cc@{\hspace{10mm}}}
    Signal yield & $\sqs=7\,\mathrm{TeV}$
    & $\sqs=8\,\mathrm{TeV}$ \\ 
    \hline
    $N_{\Y1S}$ & 326\,300 $\pm$ 638 & 747\,610 $\pm$ 969\\
    $N_{\Y2S}$ & 100\,620 $\pm$ 395 & 229\,950 $\pm$ 576\\
    $N_{\Y3S}$ & \phantom{0}57\,613 $\pm$ 312 & 129\,450 $\pm$ 459\\
  \end{tabular*}
\end{table}

Muon pairs with invariant mass in the~intervals
\mbox{$9310<m_{\mumu}<9600\mevcc$},
\mbox{$9860<m_{\mumu}<10\,155\mevcc$} and
\mbox{$10\,220<m_{\mumu}<10\,520\mevcc$}  are used as 
\Y1S, \Y2S and \Y3S candidates, respectively, when reconstructing \chib~particles. 
The~selected \ups~candidates are combined with
photons reconstructed using the~electromagnetic calorimeter
and identified using a~likelihood-based estimator, constructed from variables
that rely on calorimeter and tracking 
information~\cite{Deschamps:691634,LHCb-PAPER-2011-030,LHCb-PAPER-2012-022,LHCb-DP-2013-004}.
Candidate photon clusters
must not be  associated with the~position
of any reconstructed track extrapolated to the~calorimeter.
The~photon selection is further refined by using information
from the~\presh~and \spd~detectors.
The~photon transverse energy is required to be greater
than~600\mev.

\begin{figure*}[t!]
  \setlength{\unitlength}{1mm}
  \centering
  \begin{picture}(150,135)
  
  \newsavebox{\boxchibonesseven}
  \savebox{\boxchibonesseven}(75,45)[bl]{
    \put(0,0){
      \includegraphics*[width=75mm, height=45mm]{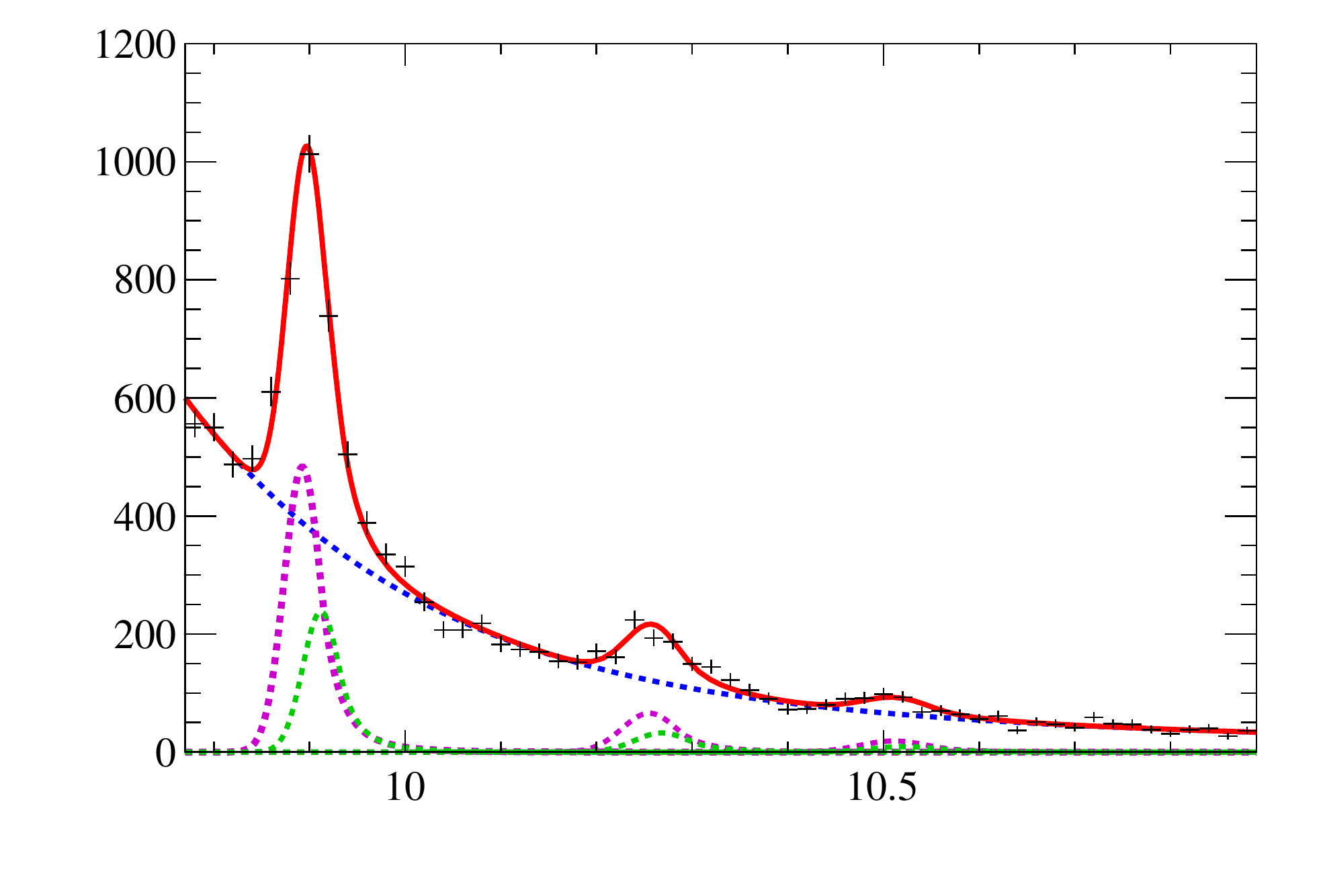}
    }
    \put(3 ,10){\scriptsize \begin{sideways}Candidates/(20\mevcc)\end{sideways}}
    \put(40,2){\scriptsize  $m_{\Y1S\g}$}
    \put(61,2){\scriptsize  $\left[\gevcc\right]$}
    \put(48,35){$\begin{array}{l}\text{LHCb} \\ \sqs=7\tev \end{array}$ }
  }

  \newsavebox{\boxchiboneseight}
  \savebox{\boxchiboneseight}(75,45)[bl]{
    \put(0,0){
      \includegraphics*[width=75mm, height=45mm]{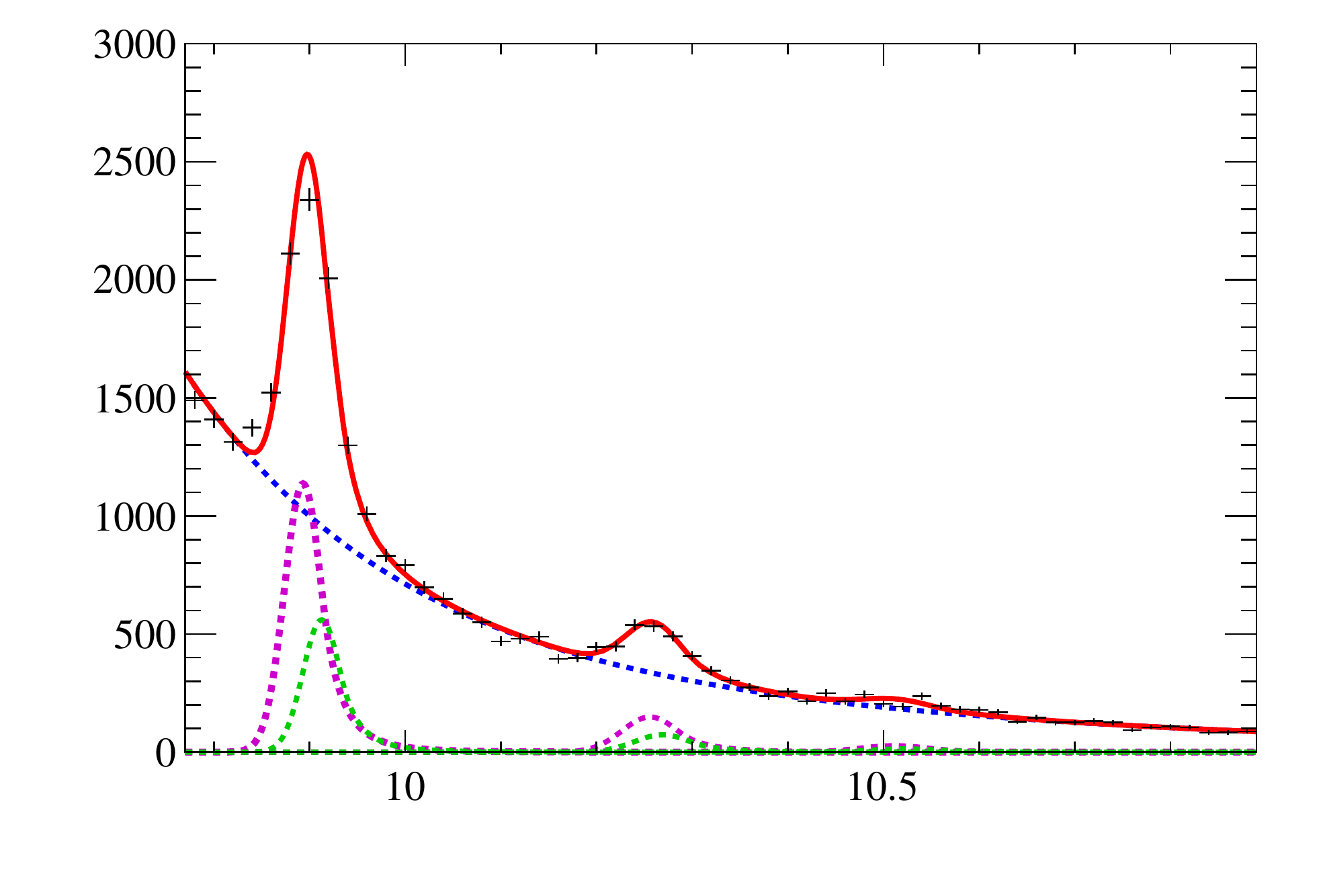}
    }
    \put(3 ,10){\scriptsize \begin{sideways}Candidates/(20\mevcc)\end{sideways}}
    \put(40,2){\scriptsize $m_{\Y1S\g}$}
    \put(61,2){\scriptsize $\left[\gevcc\right]$}
    \put(48,35){$\begin{array}{l}\text{LHCb} \\ \sqs=8\tev \end{array}$ }
  } 

  \newsavebox{\boxchibtwosseven}
  \savebox{\boxchibtwosseven}(75,45)[bl]{
    \put(0,0){
      \includegraphics*[width=75mm, height=45mm]{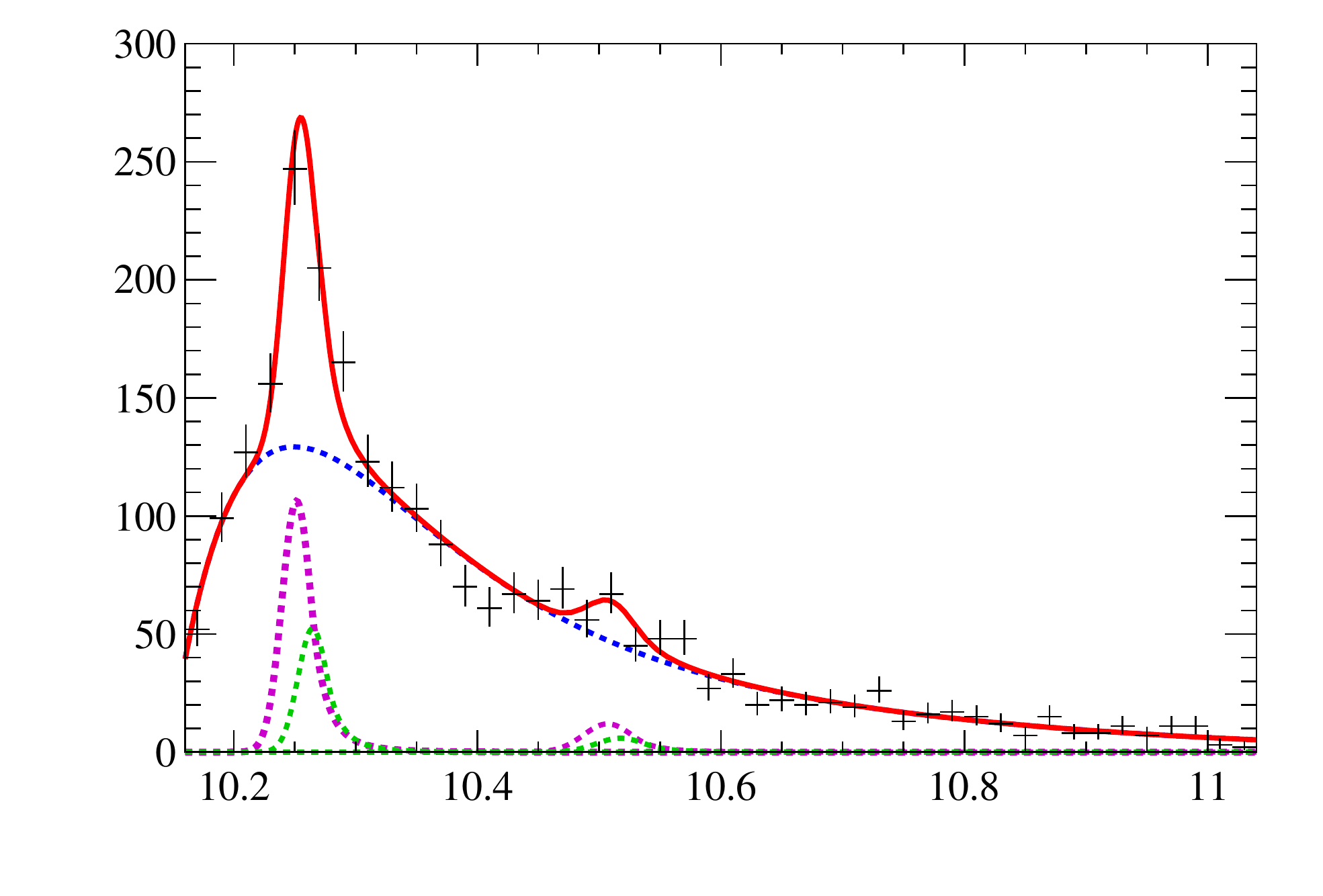}
    }
    \put(3 ,10){\scriptsize \begin{sideways}Candidates/(20\mevcc)\end{sideways}}
    \put(40,2){\scriptsize $m_{\Y2S\g}$}
    \put(61,2){\scriptsize $\left[\gevcc\right]$}
    \put(48,35){$\begin{array}{l}\text{LHCb} \\ \sqs=7\tev \end{array}$ }
  }

  \newsavebox{\boxchibtwoseight}
  \savebox{\boxchibtwoseight}(75,45)[bl]{
    \put(0,0){
      \includegraphics*[width=75mm, height=45mm]{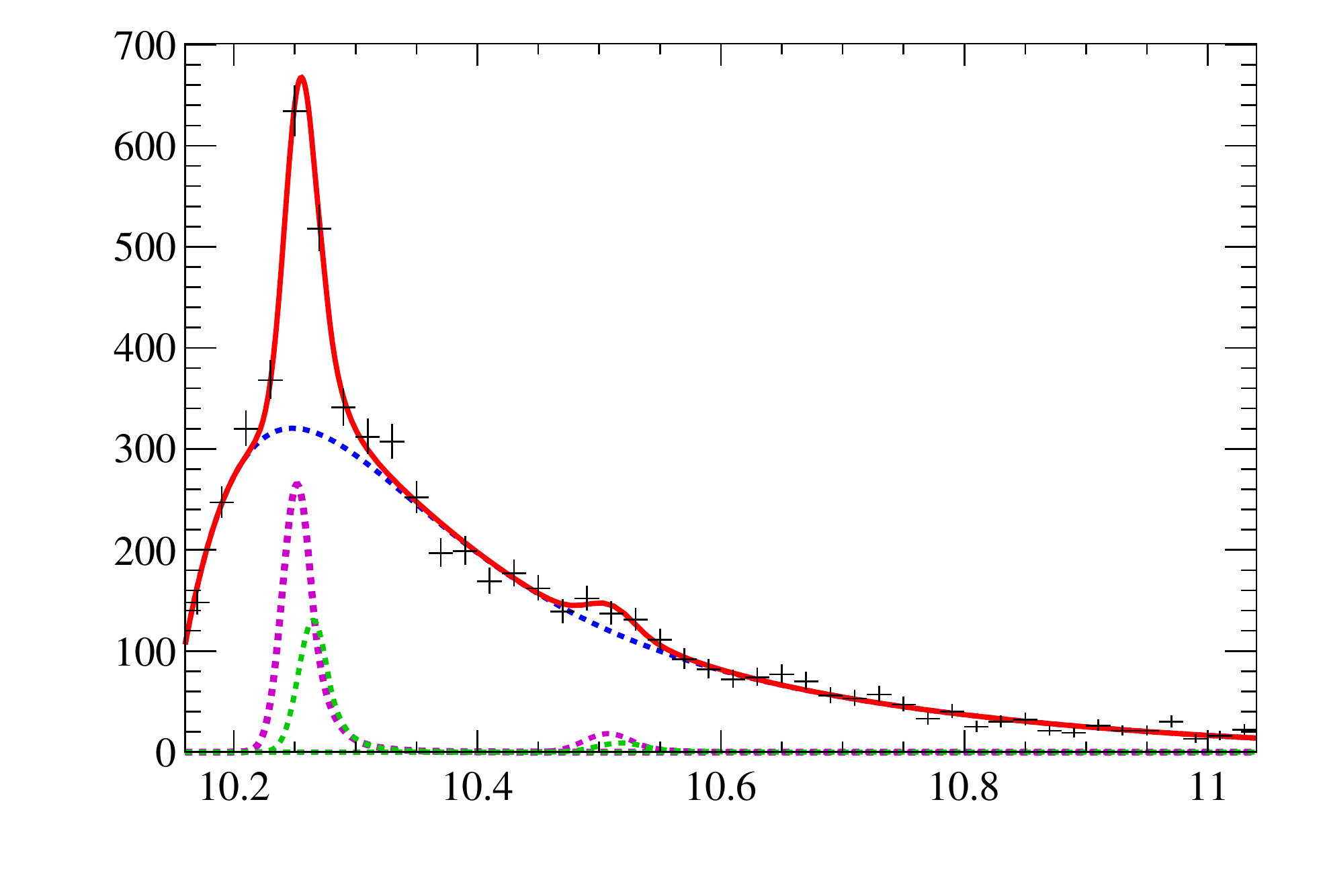}
    }
    \put(3 ,10){\scriptsize \begin{sideways}Candidates/(20\mevcc)\end{sideways}}
    \put(43,2){\scriptsize  $m_{\Y2S\g}$}
    \put(61,2){\scriptsize  $\left[\gevcc\right]$}
    \put(48,35){$\begin{array}{l}\text{LHCb} \\ \sqs=8\tev \end{array}$ }
  }

  \newsavebox{\boxchibthreesseven}
  \savebox{\boxchibthreesseven}(75,45)[bl]{
    \put(0,0){
      \includegraphics*[width=75mm, height=45mm]{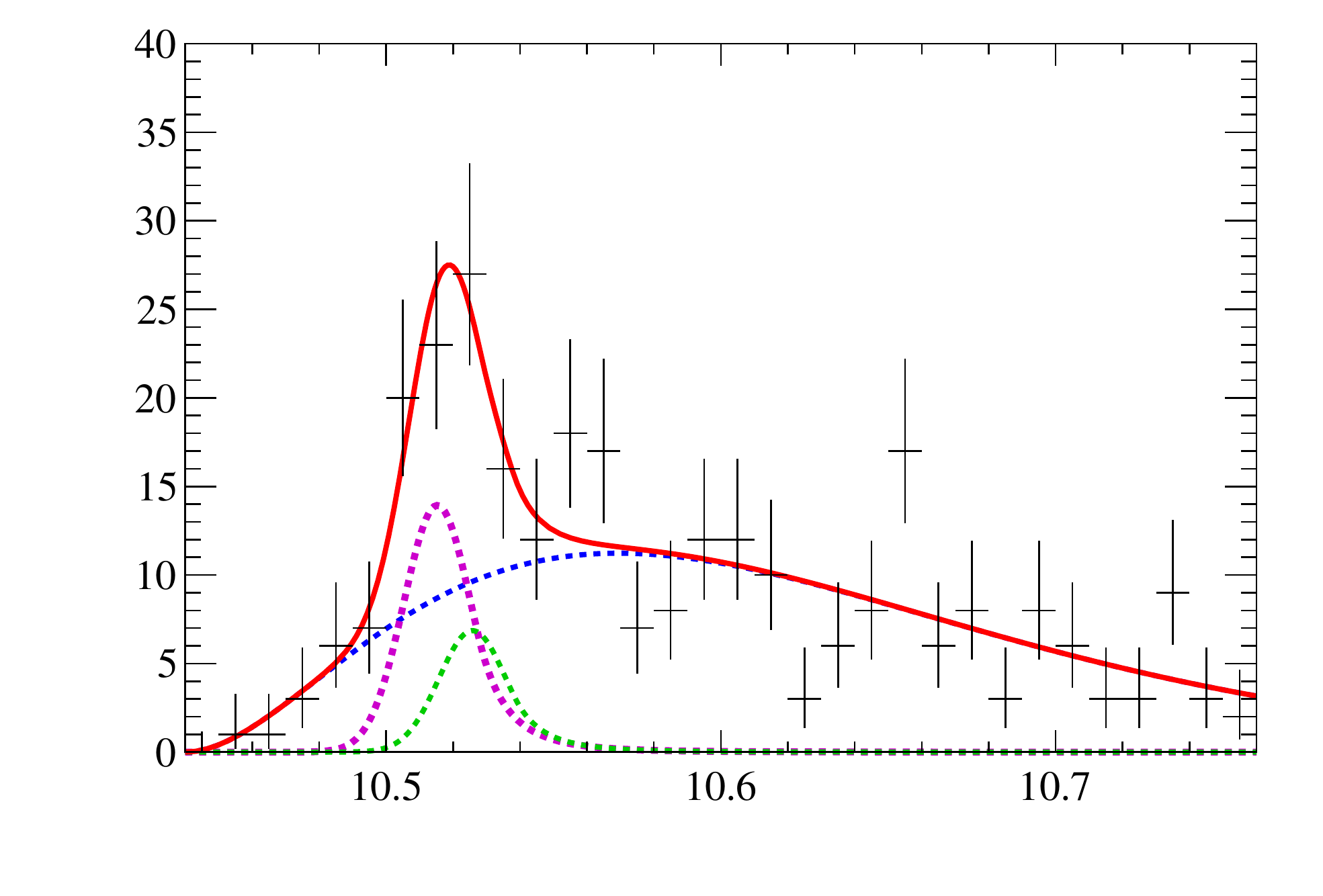}
    }
    \put(3 ,10){\scriptsize \begin{sideways}Candidates/(10\mevcc)\end{sideways}}
    \put(40,2){\scriptsize $m_{\Y3S\g}$}
    \put(61,2){\scriptsize  $\left[\gevcc\right]$}
    \put(48,35){$\begin{array}{l}\text{LHCb} \\ \sqs=7\tev \end{array}$ }
  }

  \newsavebox{\boxchibthreeseight}
  \savebox{\boxchibthreeseight}(75,45)[bl]{
    \put(0,0){
      \includegraphics*[width=75mm, height=45mm]{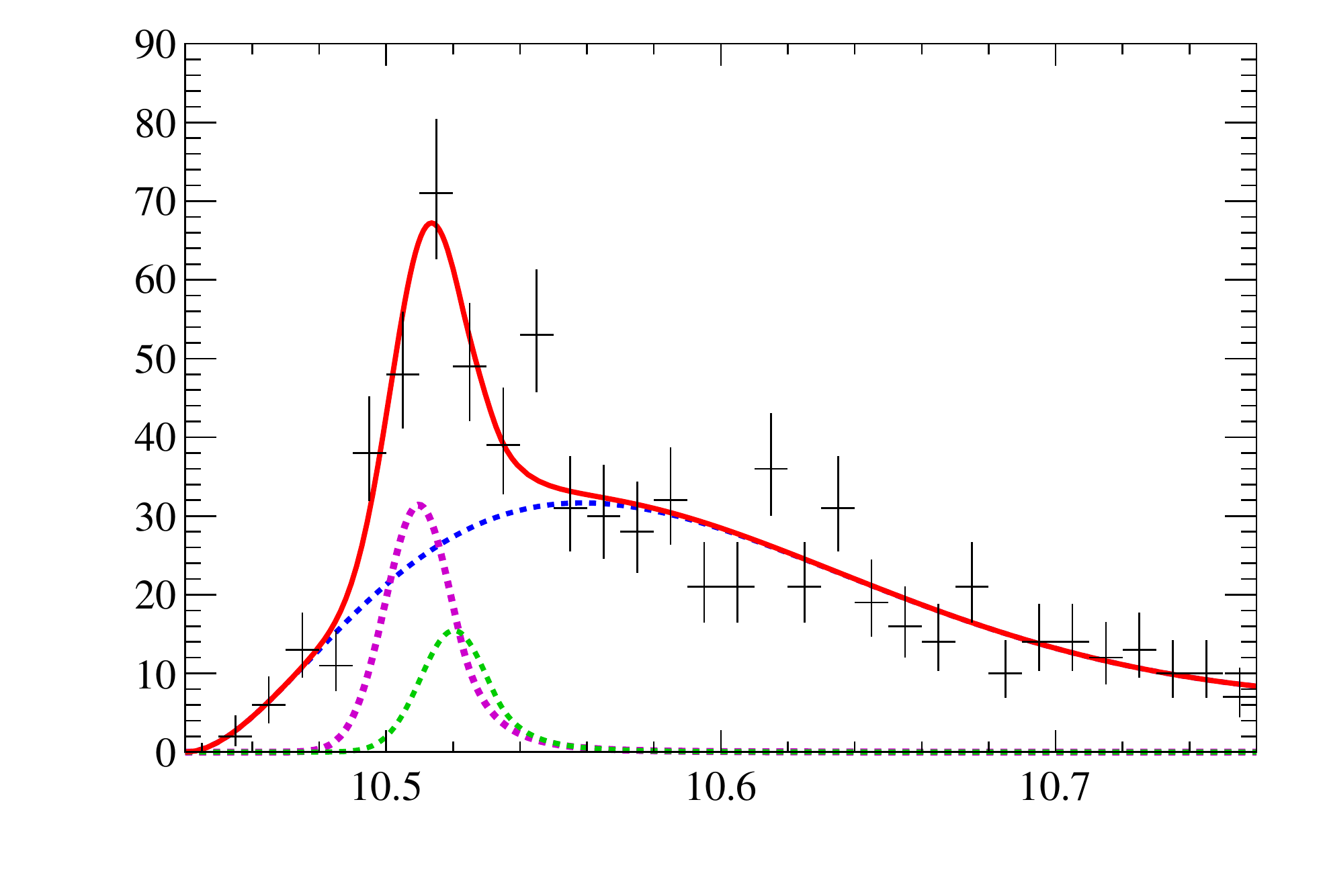}
    }
    \put(3 ,10){\scriptsize \begin{sideways}Candidates/(10\mevcc)\end{sideways}}
    \put(40,2){\scriptsize $m_{\Y3S\g}$}
    \put(61,2){\scriptsize  $\left[\gevcc\right]$}
    \put(48,35){$\begin{array}{l}\text{LHCb} \\ \sqs=8\tev \end{array}$ }
  }  

  \put(0,90){\usebox{\boxchibonesseven}}
  \put(75,90){\usebox{\boxchiboneseight}}
  \put(0,45){\usebox{\boxchibtwosseven}}
  \put(75,45){\usebox{\boxchibtwoseight}}
  \put(0,0){\usebox{\boxchibthreesseven}}
  \put(75,0){\usebox{\boxchibthreeseight}}

  \end{picture}
  \caption {\small
    Distributions of the~corrected mass $m_{\YnS\g}$ for the~selected
    \chib~candidates\,(black points) decaying into
    (top row)~\Y1S,
    (middle row)~\Y2S and 
    (bottom row)~\Y3S,
    in the~transverse momentum ranges given in the~text, for
    (left)~$\sqs=7\,\mathrm{TeV}$
    and 
    (right)~$8\,\mathrm{TeV}$~data.
    Each plot shows also the~result of the~fit
    (solid red curve), including the~background (dotted blue curve) and
    the~signal (dashed green and magenta curves) contributions.
    The~green dashed curve
    corresponds to the~\chibone~signal
    and the~magenta dashed curve to the~\chibtwo~signal.
  }
  \label{fig:chib:ups:fit:nominal}
\end{figure*}

The~\chib~signals are searched for  in the~invariant mass of $\ups\g$~combinations.
To improve the~$\YnS\g$~mass~resolution
and to~remove any residual bias,
the~corrected mass
\begin{equation}
  m_{\YnS\g} \equiv m_{\mumu\g}- \left( m_{\mumu}-m_{\YnS}\right)
\end{equation}
is used,
where $m_{\YnS}$~is the~known mass of  the~\YnS~meson~\cite{PDG2012}.
The~resolution improves by a~factor between two and four with respect
to the~one obtained 
by simply computing the invariant mass of the $\ups\g$ pair.  
The~distributions of the~corrected masses $m_{\YnS\g}$ are
shown in~Fig.~\ref{fig:chib:ups:fit:nominal} 
for \Y1S, \Y2S and \Y3S~candidates
in the~transverse momentum ranges
\mbox{$14<p_{\mathrm{T}}^{\Y1S}<40\gevc$},
\mbox{$18<p_{\mathrm{T}}^{\Y2S}<40\gevc$} and 
\mbox{$24<p_{\mathrm{T}}^{\Y3S}<40\gevc$}.

\begin{table}[t!]
  \centering
  \caption{\small Signal yields resulting from fits to
    the~corrected mass $m_{\YnS\g}$ distributions in 
    the~transverse momentum ranges \mbox{$14<p_{\mathrm{T}}^{\Y1S}<40\gevc$},
    \mbox{$18<p_{\mathrm{T}}^{\Y2S}<40\gevc$} and 
    \mbox{$24<p_{\mathrm{T}}^{\Y3S}<40\gevc$}.
    Only statistical uncertainties are shown.
  }
  \label{tab:chib:result:nominal}
  \vspace*{3mm}
  \begin{tabular*}{0.65\textwidth}{@{\hspace{10mm}}l@{\extracolsep{\fill}}cc@{\hspace{10mm}}}
    Decay mode & $\sqs=7\,\mathrm{TeV}$
    & $\sqs=8\,\mathrm{TeV}$ \\ 
    \hline
    $N_{\decay{\chibOneP}{\Y1S\g}}$    &           1908 $\pm$ 71 & 4608 $\pm$ 115\\
    $N_{\decay{\chibTwoP}{\Y1S\g}}$    & \phantom{0}390 $\pm$ 41  & \phantom{0}904 $\pm$ 68\phantom{0}\\
    $N_{\decay{\chibThreeP}{\Y1S\g}}$  & \phantom{0}133 $\pm$ 31  & \phantom{0}196 $\pm$ 50\phantom{0}\\
    $N_{\decay{\chibTwoP}{\Y2S\g}}$    & \phantom{0}265 $\pm$ 30 & \phantom{0}660 $\pm$ 46\phantom{0}\\
    $N_{\decay{\chibThreeP}{\Y2S\g}}$  & \phantom{00}48 $\pm$ 17 & \phantom{00}73 $\pm$ 26\phantom{0}\\
    $N_{\decay{\chibThreeP}{\Y3S\g}}$ & \phantom{00}56 $\pm$ 12 & \phantom{0}126 $\pm$ 20\phantom{0}
  \end{tabular*}
\end{table}

The~yields of \chibmp~mesons are determined
from an~extended maximum likelihood fit to
the~unbinned ~$m_{\YnS\g}$ distributions.  
The~fit model consists of the~sum of signal components
for all kinematically allowed \decay{\chibmp}{\YnS\g}~decays
and combinatorial background.
Neglecting a~possible contribution due to 
\decay{\Pchi_{\bquark0}\mathrm{(mP)}}{\YnS \g}~decays,
the~signal from each $\chibmp$~multiplet is parameterised
as the~sum of two overlapping Crystal Ball\,(CB) functions~\cite{Skwarnicki:1986xj}
with high-mass tails. 
The~peak positions are
separated by the~known mass-splitting between 
the~tensor and vector states
in the~\chibonep and \chibtwop~multiplets~\cite{PDG2012}.
For the~\chibthreep~multiplet
the~expected splitting of $10.5\mevcc$~\cite{Kwong:1988ae,Motyka:1997di} is used. 
The~tail parameters of the~CB~functions and the~resolutions
are fixed to the~values
determined using simulated samples.
The~yield fractions $N_{\Pchi_{\bquark 2}}/N_{\Pchi_{\bquark 1}}$ of the~tensor and vector states in each
\chibmp~multiplet are assumed to be equal to 0.5
according to expectations  from Refs.~\cite{Likhoded:2012hw,Kwong:1988ae}.
For the~\chibonep and \chibtwop cases,
this choice agrees
with direct measurements of
the~relative  productions of 
$\chibtwoOneP/\chiboneOneP$ and 
$\chibtwoTwoP/\chiboneTwoP$~\cite{CMS-PAS-BPH-13-005,LHCb-PAPER-2014-040}.
This assumption is necessary for the~determination of signal
yields, since the~\chibone and \chibtwo~states cannot be resolved given
the~limited invariant mass resolution for
the~$\YnS\g$~system. 
The~impact of this assumption is quantified as a~systematic uncertainty.
With this  parameterisation  
for the~twelve \chib~signal components,
the free parameters are
the~three masses of the~\chibone states
and the~six  overall yields  of \chibone
and \chibtwo~signals.
The~combinatorial background is parameterised as the~product
of an~exponential and polynomial functions up to the~fourth order.
The~fit results are superimposed on
Fig.~\ref{fig:chib:ups:fit:nominal} and the~signal yields
are summarized in Table~\ref{tab:chib:result:nominal}.

To perform a~precise measurement of
the~\chiboneThreeP~mass,
the~data samples collected at~\mbox{$\sqs=7$}~and 8\tev are combined.
A~fit to the~combined sample of ${\decay{\chibThreeP}{\Y3S\g}}$~decays gives
\begin{equation*} 
m_{\chiboneThreeP} = 10\,511.3 \pm 1.7\mevcc,
\end{equation*}
where the~uncertainty is statistical only.

For the~determination of the~\chib~signal  yields  in $p_{\mathrm{T}}^{\ups}$~bins,
the~masses
of the~\chibone~states in the~fits are fixed to the~values obtained in the~fits to
the~full \pt~ranges.
For each $p_{\mathrm{T}}^{\ups}$ bin the~fractions
$\mathcal{R}^{\chibmp}_{\YnS}$, defined by Eq.~\eqref{eq:r}, are calculated
separately for \sqs=7~and 8\tev data  samples
as
\begin{equation}
  \mathcal{R}^{\chibmp}_{\YnS} =
  \dfrac{ N_{\chibmp} } { N_{\YnS}}
  \times
  \dfrac
      { \varepsilon_{\YnS}}
      { \varepsilon_{\chibmp} },
      \label{eq:calc}
\end{equation}
where
$\varepsilon_{\chibmp}$ and $\varepsilon_{\YnS}$ denote
the~total efficiencies,  
and $N_{\chibmp}$ and $N_{\YnS}$ are
the~fitted yields for the~\chibmp and \YnS states
for the~respective 
$p_{\mathrm{T}}^{\ups}$~bin.
The~ratio of the~efficiencies
$\varepsilon_{\chibmp}$ and $\varepsilon_{\YnS}$
is largely determined by 
the~reconstruction efficiency for photons
from \chib~decays.
It is close to 25\% for
\chib~mesons with transverse momentum larger than~20\gevc,
and it drops to approximately 10\% for the~lowest \pt considered in this analysis.
The~dominant sources of inefficiency
are the~geometrical acceptance of
the~electromagnetic calorimeter,
photon conversions in the~detector material,
the~accidental overlap of clusters in the~\ecal 
and the~selection requirement on the~photon transverse energy.
The~measurements are performed in
six bins of $p_{\mathrm{T}}^{\Y1S}$  in the~range \mbox{$6<p_{\mathrm{T}}^{\Y1S}<40\gevc$}, 
five bins of $p_{\mathrm{T}}^{\Y2S}$  in the~range \mbox{$18<p_{\mathrm{T}}^{\Y1S}<40\gevc$} and
two bins of $p_{\mathrm{T}}^{\Y3S}$  in the~range \mbox{$24<p_{\mathrm{T}}^{\Y1S}<40\gevc$}.

\section{Systematic uncertainties}
\label{sec:Systematics}

The~systematic uncertainties on the~fractions $\mathcal{R}^{\chibmp}_{\YnS}$,
calculated using Eq.~\eqref{eq:calc},
are related to the~determination of
the~signal yields and the~evaluation of the~efficiency ratios. 
The~main contributions to the~former are due to fit modeling, whereas the 
photon reconstruction efficiency and the knowledge of the initial state polarization 
dominate the uncertainty on the ratios of efficiencies
$\varepsilon_{\chibmp}/\varepsilon_{\YnS}$. 
The contributions due to other effects largely cancel in these ratios.

Based on studies from Refs.~\cite{LHCb-PAPER-2011-036,LHCb-PAPER-2012-015,LHCb-PAPER-2013-016,LHCb-PAPER-2013-066}
the~systematic uncertainty associated with the~\ups~signal yields determination
is taken to be 0.7\% for all $p_{\mathrm{T}}^{\ups}$~bins.

In the~\chib~fit model several sources of uncertainty  are taken into account.
The~yield ratio $N(\chibtwo)/N(\chibone)$,
which is fixed in the~fit to be 0.5
as~predicted by theory, is varied from 0.3 to 1.0.
These limits are obtained by following the~prescription of 
Ref.~\cite{Likhoded:2012hw}, where the~experimentally measured
cross-section ratio of \chic~mesons is rescaled to predict the~corresponding ratio
for \chib mesons. The~ratio of cross-sections is then converted to a~ratio of yields 
by taking into account the~\chibone and \chibtwo radiative branching fractions and 
reconstruction efficiencies.
For the~\chibonep and the~\chibtwop~mesons, the~variation obtained 
agrees within uncertainties with the~direct measurements of
relative  productions of $\chibtwoOneP$~and $\chiboneOneP$~mesons
and $\chibtwoTwoP$ and $\chiboneTwoP$~mesons~\cite{LHCb-PAPER-2014-040}. 
The~corresponding systematic uncertainty on $\mathcal{R}^{\chibmp}_{\YnS}$ 
varies between 0.1\% and 15\% across $p_{\mathrm{T}}^{\ups}$~bins.
The~systematic uncertainty due to a~slight
dependence of the~mass fit results on $p_{\mathrm{T}}^{\ups}$
is estimated 
by taking the~minimum and the~maximum values of the~$\chibone$~masses,
repeating  the~fit and taking the~maximum difference in the~yields. 
The~assigned uncertainty varies between 0.3\% and 20\%
for various $p_{\mathrm{T}}^{\ups}$~bins.
The~smaller values  corresponds to the~low-Q transitions:
\mbox{\decay{\chibonep}{\Y1S\g}},
\mbox{\decay{\chibtwop}{\Y2S\g}} and 
\mbox{\decay{\chibthreep}{\Y3S\g}}.
To~assess the~systematic uncertainty related to possible
mismodelling of  the~mass~resolution, the~mass  resolution
is varied by $\pm10\%$~around the~values obtained using simulated
samples, and the~difference between the~obtained $\mathcal{R}^{\chibmp}_{\YnS}$
is treated as the~corresponding systematic uncertainty. 
The~maximum deviation in the~results obtained from varying 
by $\pm1$~the~order of the~polynomial function
used in the~fit model
to describe the~combinatorial background,
is assigned as the~systematic uncertainty associated with
the~background parameterisation.
For~the~\chibthreep case, a~systematic uncertainty stems from
the~assumption on   the~mass splitting between
\chibtwoThreeP and \chiboneThreeP~states.
This parameter is varied in the~range between 9~and 12\mevcc.
The~obtained uncertainty for $\mathcal{R}^{\chibthreep}_{\Y3S}$ is found
to be much smaller than the~one obtained for $\mathcal{R}^{\chibthreep}_{\Y1S}$ and
$\mathcal{R}^{\chibthreep}_{\Y2S}$.
The~assigned uncertainty on  $\mathcal{R}^{\chibthreep}_{\YnS}$
varies between 0.1\% and 2\%.

\begin{table}[!t]
  \centering
  \caption{\small
    Summary of the~relative systematic uncertainties
    for the~fractions $\mathcal{R}^{\chibmp}_{\YnS}$.
  }
  \label{tab:syst}
  \vspace*{3mm}
  \begin{tabular*}{0.60\textwidth}{@{\hspace{10mm}}l@{\extracolsep{\fill}}c@{\hspace{10mm}}}
    Source  & Uncertainty~$\left[\%\right]$
    \\
    \hline
    \ups~fit model                         &  0.7 
    \\
    \chib~fit model                        &
    \\
    ~~$\chibone/\chibtwo$~ratio            & 0.1 -- 15
    \\
    ~~\chibone~mass variation              & 0.3 -- 20
    \\
    ~~\chib~mass resolution                & 2.0 -- 12
    \\
    ~~background model                     & 2.0 -- 10
    \\
    ~~$m_{\chibtwoThreeP}-m_{\chiboneThreeP}$      & 0.1 -- \phantom{0}2
    \\
    \g~reconstruction                      &  3.0
    \\
    \chib~polarization                     & 0.9 -- \phantom{0}9
  \end{tabular*}
\end{table}

The~uncertainty due to possible imperfections in the~simulation in the~determination of
the~photon reconstruction efficiency is studied by comparing the~relative yields
between data and simulation for  $\Bu\to\jpsi\mathrm{K}^{*+}$ and  $\Bu\to\jpsi\mathrm{K}^{+}$ decays,
where the~$\mathrm{K}^{*+}$~meson is reconstructed using
the~$\mathrm{K}^+\piz$ final 
state~\cite{LHCb-PAPER-2012-015,LHCb-PAPER-2012-022,LHCb-PAPER-2012-053,LHCb-PAPER-2013-024,LHCb-PAPER-2014-008}.
According to these studies, a~systematic uncertainty of 3\% is assigned
for photons
in the~kinematical range considered in this analysis.
This uncertainty is dominated by 
the~knowledge of the~ratio
of the~branching fractions for 
$\Bu\to\jpsi\mathrm{K}^{*+}$ and  $\Bu\to\jpsi\mathrm{K}^{+}$~decays.

Another source of systematic uncertainty is associated with
the~unknown polarization of \chib~and \ups~states.
The~polarization of \ups~mesons for $p_{\mathrm{T}}^{\ups}>10\gevc$ and
in the~central rapidity region $\left|y^{\ups}\right|<1.2$
has been found to be small by the~CMS collaboration~\cite{Chatrchyan:2012woa}.
Therefore in this paper we assume zero polarization
of \ups~mesons and no systematic uncertainty is assigned due to this effect.
The~systematic uncertainty related to the~unknown
polarization of \chib~mesons was estimated following Refs.~\cite{Abt:2008ed,LHCb-PAPER-2013-028}.
For each $p_{\mathrm{T}}^{\ups}$~bin, the~ratios of efficiencies
$\varepsilon_{\chibone}/\varepsilon_{\ups}$~and
$\varepsilon_{\chibtwo}/\varepsilon_{\ups}$~are recomputed using
various possible
polarizations
scenarios
for~\chibone~and \chibtwo~mesons.
The~maximum deviation of the~efficiency ratio
with respect to the~one obtained with unpolarized production of 
\chibone and \chibtwo~states is taken as the~systematic uncertainty.
The~assigned uncertainty on  $\mathcal{R}^{\chibmp}_{\YnS}$
varies between 0.9\%~and 9\%~for various $p_{\mathrm{T}}^{\ups}$~bins.

Systematic uncertainties due to
external experimental inputs, \eg 
the~$\ups$~mass or the~mass splitting of \chibonep and \chibtwop~multiplets,  
are negligible.
The~systematic uncertainties
on the $\mathcal{R}^{\chibmp}_{\YnS}$~measurements are summarized  in Table~\ref{tab:syst}.

Systematic uncertainties on the~measurement of
the~\chiboneThreeP mass are due to the~\ecal~energy scale,
the~fit model and the \Y3S~mass~\cite{PDG2012}.
The~firs of these is studied by comparing the~reconstructed invariant mass of
photons in \decay{\piz}{\g\g}~decays with
the~known mass of the~neutral
pion~\cite{Kali,Dasha_disser,CalibPET},  
which gives an uncertainty of 1.0\mevcc in \decay{\chibthreep}{\Y3S\g}~decays.
The~effects of possible mismodelling of  the~mass~resolution
and background models are found to be
0.8\mevcc and 0.3\mevcc, respectively.
Other significant contributions to the~systematic uncertainty 
are related to the~assumptions on~$N(\chibtwo)/N(\chibone)$,
and to the~mass splitting  between \chib~multiplet components.
The~effect of the~unknown value for the~mass-splitting
is tested by varying \mbox{$m_{\chibtwoThreeP}-m_{\chiboneThreeP}$}
in the~fit
in a~range between  9 and 12\mevcc,
preferred by theory~\cite{Kwong:1988ae,Motyka:1997di};
the~obtained deviation of 0.4\mevcc is
assigned as the~corresponding systematic uncertainty.
The~\chiboneThreeP mass exhibits a~linear dependence on the~assumed
fraction of \chibone~decays and varies from $10\,509$ to $10\,513$\mevcc,
when the~$\chibtwo/\chibone$~yield ratio changes from 0.3 to 1.0.
The~determination of the~\chiboneThreeP~mass
is further checked using the~large $\chibonep\to\Y1S\g$~signal,
where the~measured \chiboneOneP~mass agrees with
the~known \chiboneOneP~mass~\cite{PDG2012} to better
than 0.5\mevc, separately for \mbox{$\sqs=7$}~and 8\tev~data.
No additional systematic uncertainty is assigned. 
The~systematic uncertainties
on the~\chiboneThreeP~mass measurement
are summarized  in Table~\ref{tab:mass}.

\begin{table}[t!]
  \centering
  \caption{\small
    Summary of systematic uncertainties
    for $m_{\chiboneThreeP}$. 
  }
  \label{tab:mass}
  \vspace*{3mm}
  \begin{tabular*}{0.60\textwidth}{@{\hspace{10mm}}l@{\extracolsep{\fill}}c@{\hspace{10mm}}}
    Source                                 & Uncertainty $\left[\!\mevcc\right]$
    \\
    \hline
    \chib~fit model                        &
    \\
    ~~\chib~mass resolution                &  0.8
    \\
    ~~background model                     &  0.3
    \\
    ~~$m_{\chibtwoThreeP}-m_{\chiboneThreeP}$      &  0.4
    \\
    ~~$\chibone/\chibtwo$~ratio            & $2.0$
    \\
    \ecal~energy scale                     &  1.0
    \\
    \Y3S mass uncertainty                   &  0.5 
  \end{tabular*}
\end{table}

\section{Results and conclusion}
\label{sec:Conclusion}

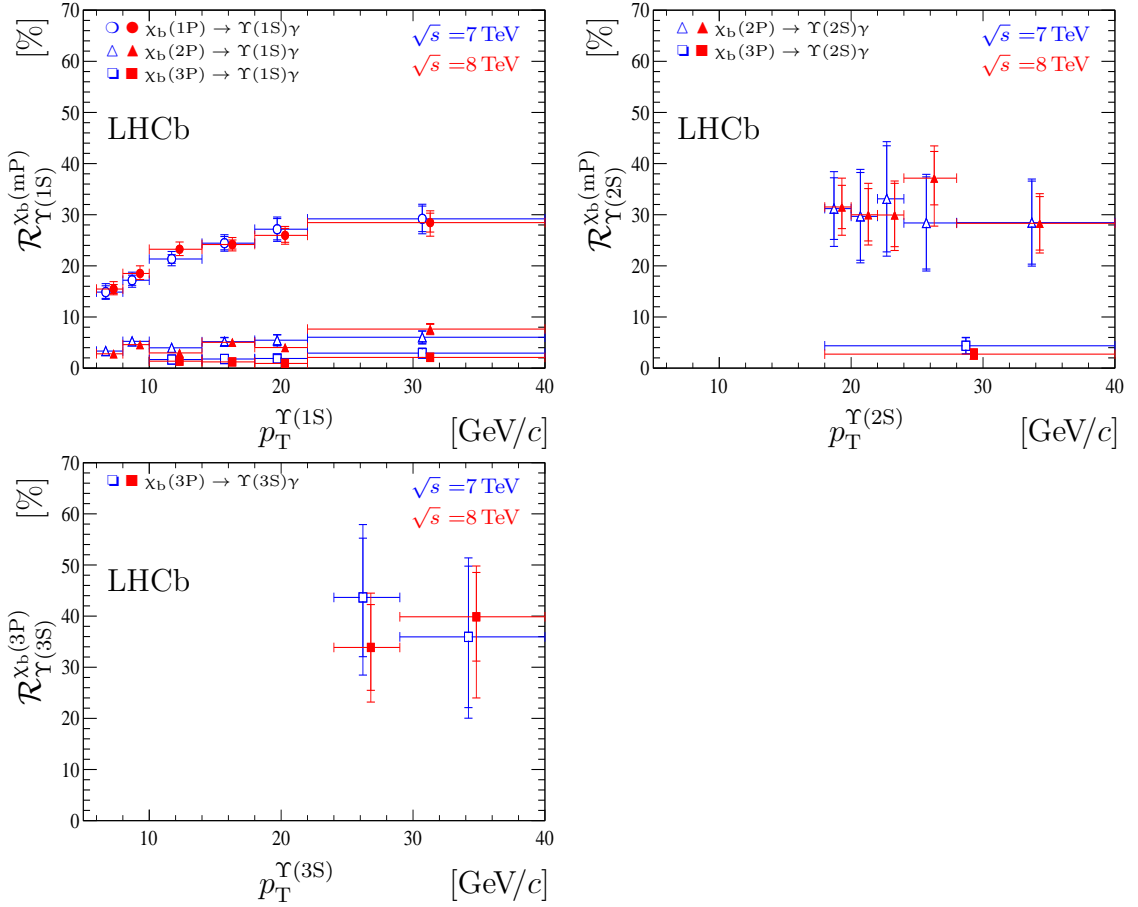
\begin{figure*}[t!]
  \setlength{\unitlength}{1mm}
  \centering
  \begin{picture}(150,120)
    \put(0,0){
      \includegraphics*[width=75mm, height=60mm]{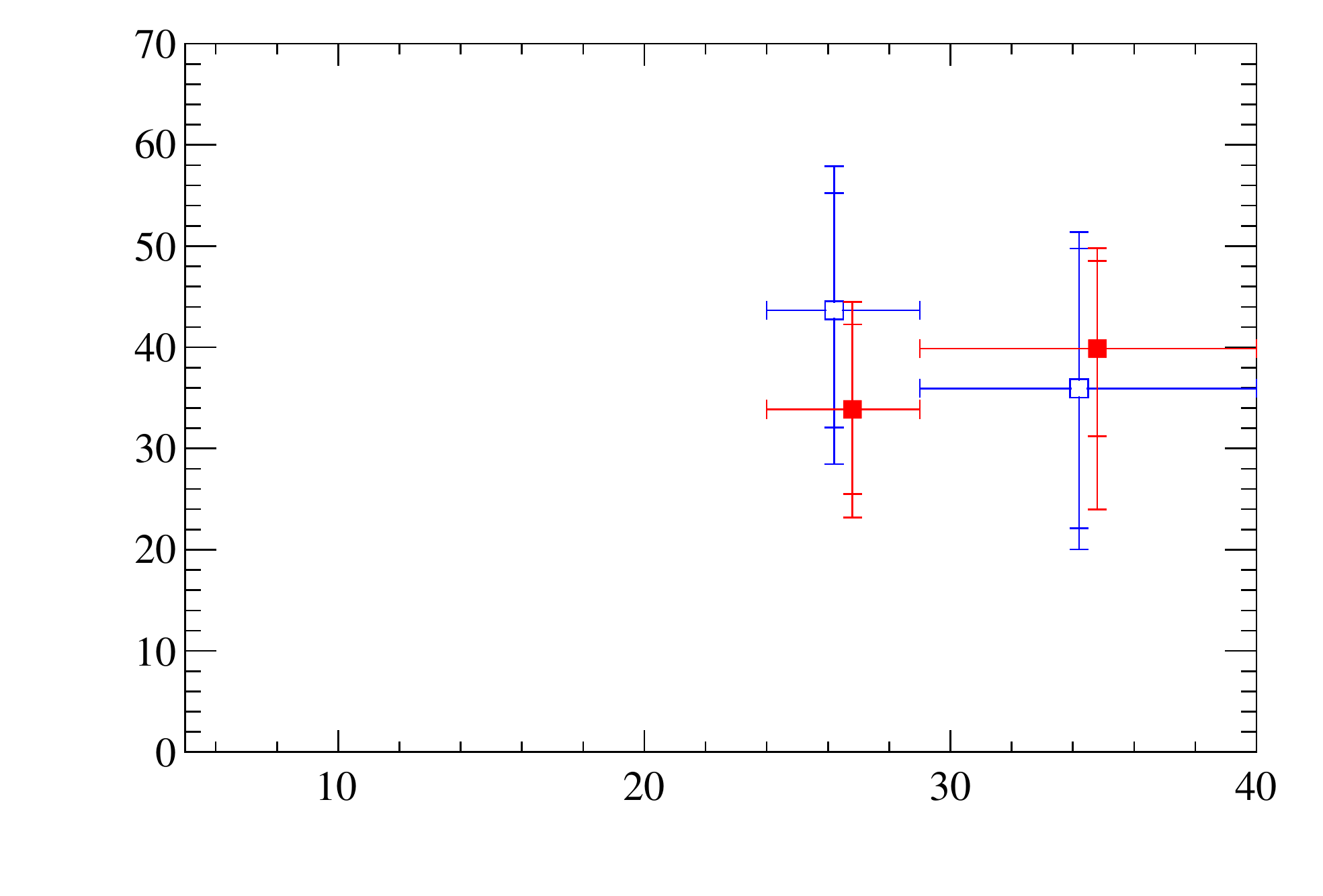}
    }
    \put(2,25){\begin{sideways}$\mathcal{R}^{\chibthreep}_{\Y3S}~~~~~~~~\left[\%\right]$ \end{sideways}}
    \put(35,0){$p_{\mathrm{T}}^{\Y3S}$}
    \put(60,0){$\left[\!\gevc\right]$}
    
    \put(55,53){\scriptsize \textcolor{blue}{\sqs=7\tev}}
    \put(55,49){\scriptsize \textcolor{red}{\sqs=8\tev}}
    
    \put(15,54){\tiny \textcolor{blue}{\ding{111}} \textcolor{red}{\ding{110}}  $\chibThreeP \to \Y3S \g$}
      
    \put(15,40){\lhcb}
    \put(0,60){
      \includegraphics*[width=75mm, height=60mm]{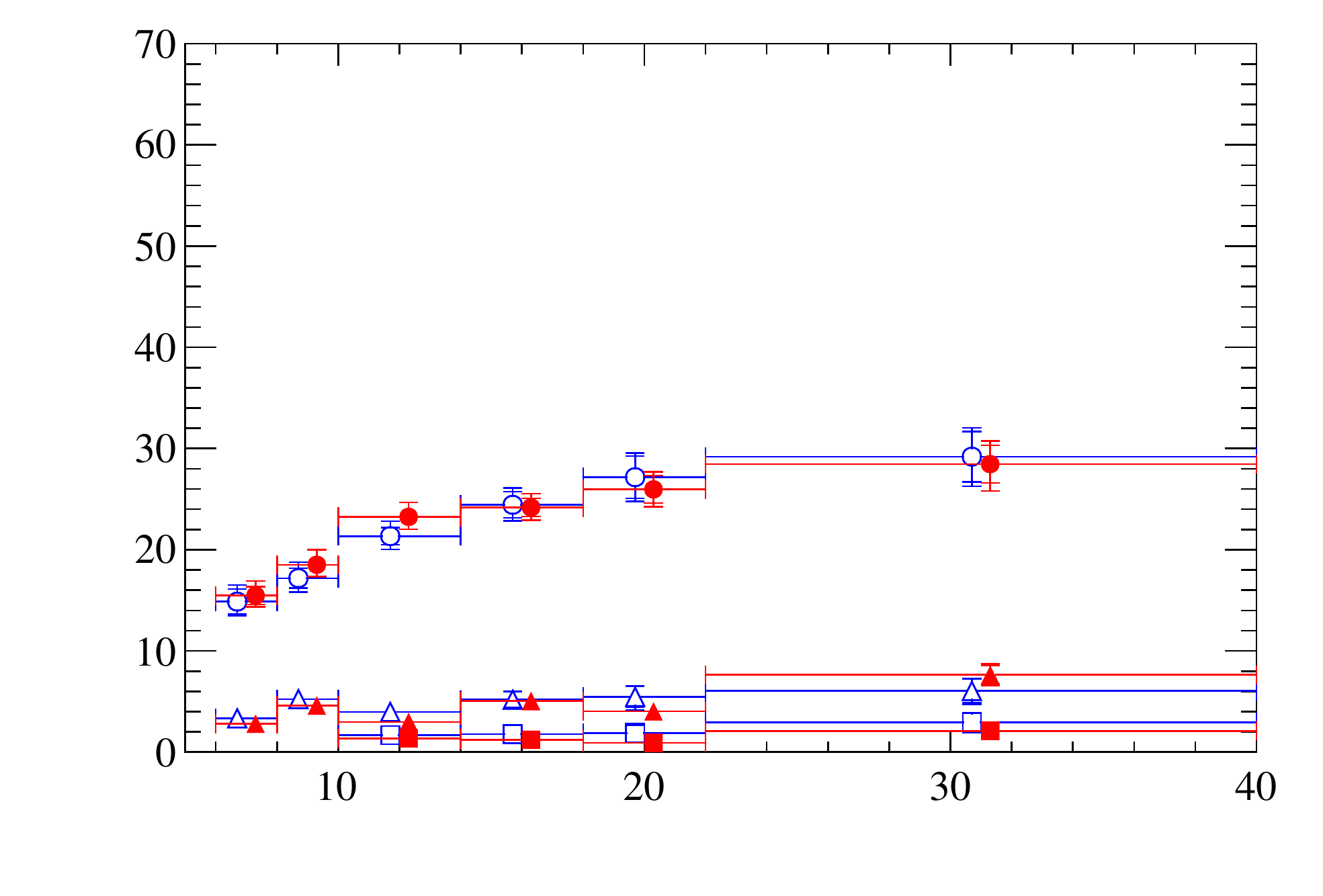}
    }
    \put(2,85){\begin{sideways}$\mathcal{R}^{\chibmp}_{\Y1S}~~~~~~~~\left[\%\right]$ \end{sideways}}
    \put(35,60){$p_{\mathrm{T}}^{\Y1S}$}
    \put(60,60){$\left[\!\gevc\right]$}
    
    \put(55,113){\scriptsize \textcolor{blue}{\sqs=7\tev}}
    \put(55,109){\scriptsize \textcolor{red}{\sqs=8\tev}}
    
    \put(14.8,114){\tiny \textcolor{blue}{\ding{109}} \textcolor{red}{\ding{108}} $\chibOneP \to \Y1S \g$}
    
    \put(14.8,111){\scalebox{0.6}{\textcolor{blue}{$\vartriangle$} \textcolor{red}{$\blacktriangle$}}\,\tiny$\chibTwoP \to \Y1S \g$}
    
    \put(15,108){\tiny \textcolor{blue}{\ding{111}} \textcolor{red}{\ding{110}} $\chibThreeP \to \Y1S \g$}    
    
    \put(15,100){\lhcb}

    \put(75,60){
      \includegraphics*[width=75mm, height=60mm]{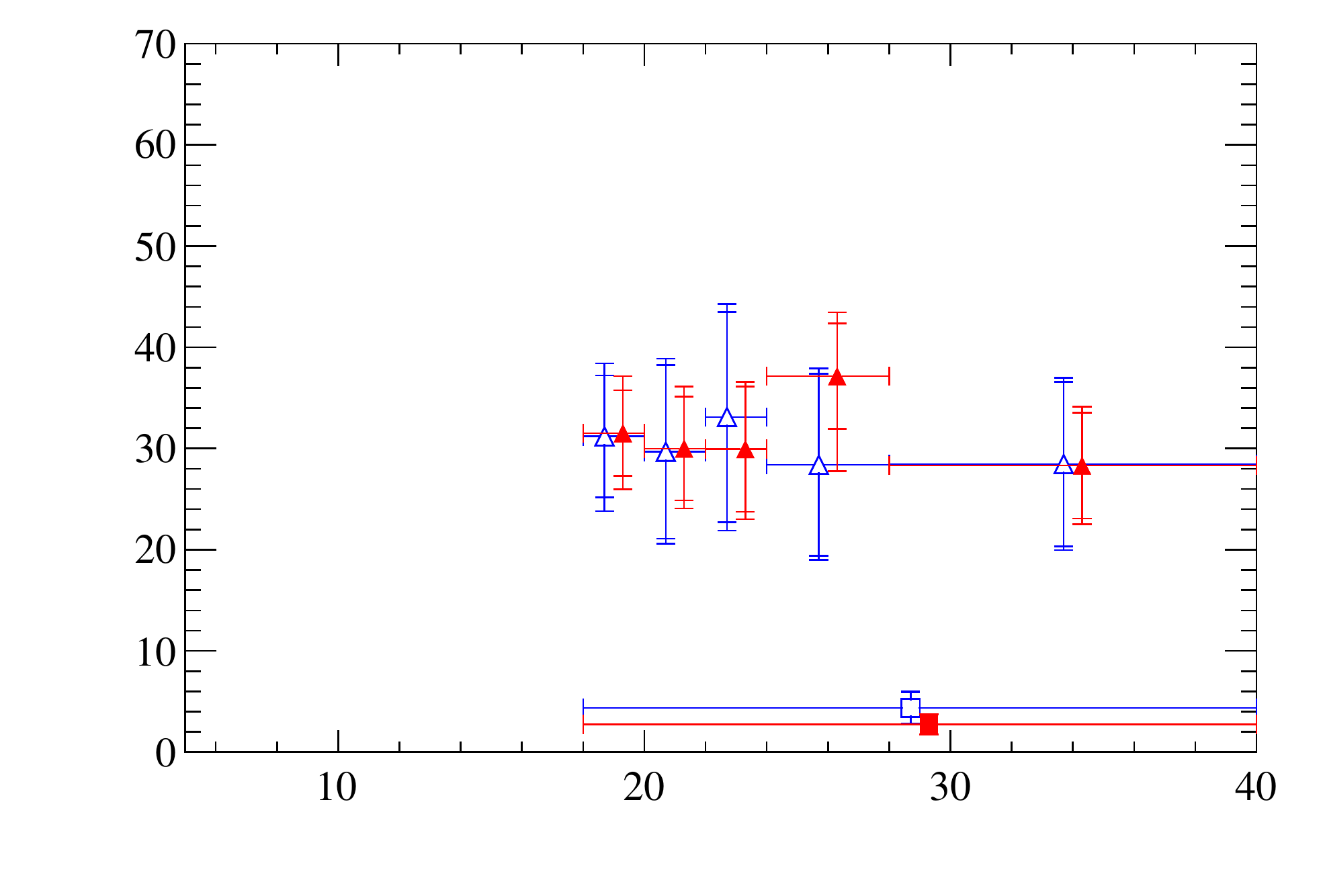}
    }
    \put(77,85){\begin{sideways}$\mathcal{R}^{\chibmp}_{\Y2S}~~~~~~~~\left[\%\right]$ \end{sideways}}
    \put(110,60){$p_{\mathrm{T}}^{\Y2S}$}
    \put(135,60){$\left[\!\gevc\right]$}
        
    \put(130,113){\scriptsize \textcolor{blue}{\sqs=7\tev}}
    \put(130,109){\scriptsize \textcolor{red}{\sqs=8\tev}}

    \put(89.8,114){\scalebox{0.6}{\textcolor{blue}{$\vartriangle$} \textcolor{red}{$\blacktriangle$}}\,\tiny$\chibTwoP \to \Y2S \g$}
    
    \put(90,111){\tiny \textcolor{blue}{\ding{111}} \textcolor{red}{\ding{110}} $\chibThreeP \to \Y2S \g$}  

    \put(90,100){\lhcb}

 \end{picture}
 \caption{\small
        Fractions \Rmn as functions of $p_{\mathrm{T}}^{\ups}$.
        Points with blue~open\,(red solid) symbols
        correspond to data collected at $\sqs=7(8)\,\mathrm{TeV}$, respectively.
        For better visualization the~data points are slightly displaced
        from the~bin centres.
        The~inner error bars represent statistical uncertainties,
        while the~outer error bars indicate statistical and systematic uncertainties added in quadrature. }
 \label{fig:frac}
\end{figure*}

The measured fractions \Rmn are presented in Fig.~\ref{fig:frac} and Tables~\ref{tab:chib:resultseven} and~\ref{tab:chib:resulteight}.
The results are dominated by the statistical uncertainties, and show no dependence on the \proton\proton~collision energy.
A~measurement of the~$\mathcal{R}^{\chibthreep}_{\Y3S}$~fraction is performed for the first time.
The large value of this fraction impacts the~interpretation
of experimental data on \ups~production and polarization.
When data on \ups~production and polarization are compared with theory predictions,
as well as when different theory predictions are compared among themselves, it is often implicitly assumed
that the~fraction of \Y3S~mesons produced by feed~down from higher states is small.
The~large measured value of  $\mathcal{R}^{\chibthreep}_{\Y3S}$
indicates that these assumptions need to be revisited.

\begin{table*}[t]
  \centering
  \caption{\small
    Fractions \Rmn in bins of~$p_{\mathrm{T}}^{\ups}$,
    measured for data collected at $\sqs=7\,\mathrm{TeV}$.
    The~first block corresponds to
    $\mathcal{R}^{\chibmp}_{\Y1S}$,
    the second 
    to~$\mathcal{R}^{\chibmp}_{\Y2S}$
    and the third to~$\mathcal{R}^{\chibmp}_{\Y3S}$.
    The first uncertainty is statistical and the second systematic.
  }
  \label{tab:chib:resultseven}
  \vspace*{3mm}
  \begin{tabular*}{0.95\textwidth}{@{\hspace{3mm}}c@{\extracolsep{\fill}}cccc@{\hspace{3mm}}}
    & $p_{\mathrm{T}}^{\ups}~\left[\!\gevc\right]$
    & $\mathcal{R}^{\chibonep}_{\YnS}$
    & $\mathcal{R}^{\chibtwop}_{\YnS}$
    & $\mathcal{R}^{\chibthreep}_{\YnS}$
    \\
    \hline
    \multirow{6}{*}{\Y1S} 
    & \phantom{0}6 -- \phantom{0}8 & $14.8\pm1.2 \pm 1.3 $ & $3.3\pm0.6 \pm 0.2   $ & \\
    & \phantom{0}8 -- 10           & $17.2\pm1.0 \pm 1.4 $ & $5.2\pm0.6 \pm 0.3   $ & \\
    & 10 -- 14                     & $21.3\pm0.8 \pm 1.4 $ & $4.0\pm0.5 \pm 0.3   $ & $1.7\pm0.5 \pm 0.1 $\\
    & 14 -- 18                     & $24.4\pm1.3 \pm 1.2 $ & $5.2\pm0.8 \pm 0.4   $ & $1.8\pm0.6 \pm 0.2 $\\
    & 18 -- 22                     & $27.2\pm2.1 \pm 2.1 $ & $5.5\pm1.0 \,\, ^{+\,\,\,0.4}_{-\,\,\,1.0}$ & $1.9\pm0.7 \pm 0.3 $\\
    & 22 -- 40                     & $29.2\pm2.5 \pm 1.7 $ & $6.0\pm1.2 \,\, ^{+\,\,\,0.4}_{-\,\,\,0.7}$ & $2.9\pm1.0 \pm 0.4 $\\
    \hline
    \multirow{6}{*}{\Y2S} 
    & 18 -- 20  &  & $31\phantom{.}\pm6\phantom{.0}   \pm 4\phantom{.0} $ & \\
    & 20 -- 22  &  & $30\phantom{.}\pm9\phantom{.0}   \pm 3\phantom{.0} $ & \\
    & 22 -- 24  &  & $33\phantom{.}\pm10\phantom{.}   \pm 5\phantom{.0} $ & \\
    & 24 -- 28  &  & $28\phantom{.}\pm9\phantom{.0}   \pm 3\phantom{.0} $ & \\
    & 28 -- 40  &  & $29\phantom{.}\pm8\phantom{.0}   \pm 3\phantom{.0} $ & \\
    & 18 -- 40  &  &                                                      & $4.4\pm1.6 \pm0.5 $\\
    \hline
    \multirow{2}{*}{\Y3S} 
    & 24 -- 29 &  &  & $\phantom{.}44\pm12\phantom{.} \pm 10\phantom{.} $\\
    & 29 -- 40 &  &  & $\phantom{.}36\pm14\phantom{.} \pm 8\phantom{0.} $\\
  \end{tabular*}
\end{table*}

\begin{table*}[t]
  \centering
  \caption{\small
    Fractions \Rmn in bins of~$p_{\mathrm{T}}^{\ups}$,
    measured for data collected at $\sqs=8\,\mathrm{TeV}$.
    The~first block corresponds to
    $\mathcal{R}^{\chibmp}_{\Y1S}$,
    the second to~$\mathcal{R}^{\chibmp}_{\Y2S}$
    and the third to~$\mathcal{R}^{\chibmp}_{\Y3S}$.
    The first uncertainty is statistical and the second systematic.
  }
  \label{tab:chib:resulteight}
  \vspace*{3mm}
  \begin{tabular*}{0.95\textwidth}{@{\hspace{3mm}}c@{\extracolsep{\fill}}cccc@{\hspace{3mm}}}
    & $p_{\mathrm{T}}^{\ups}~\left[\!\gevc\right]$
    & $\mathcal{R}^{\chibonep}_{\YnS}$
    & $\mathcal{R}^{\chibtwop}_{\YnS}$
    & $\mathcal{R}^{\chibthreep}_{\YnS}$
    \\
    \hline
    \multirow{6}{*}{\Y1S} 
    & \phantom{0}6 -- \phantom{0}8         & $15.5\pm0.9 \pm 1.3 $ & $2.8\pm0.5 \pm 0.2 $ & \\
    & \phantom{0}8 -- 10                   & $18.5\pm0.7 \pm 1.5 $ & $4.6\pm0.4 \pm 0.3 $ & \\
    & 10 -- 14                             & $23.2\pm0.6 \pm 1.4 $ & $3.0\pm0.4 \pm 0.2 $ & $1.4\pm0.4 \pm 0.1 $\\
    & 14 -- 18                             & $24.2\pm0.9 \pm 1.2 $ & $5.0\pm0.5 \pm 0.3 $ & $1.2\pm0.4 \pm 0.1 $\\
    & 18 -- 22                             & $26.0\pm1.4 \pm 1.2 $ & $4.0\pm0.7 \pm 0.3 $ & $0.9\pm0.5 \pm 0.1 $\\
    & 22 -- 40                             & $28.5\pm1.8 \pm 2.1 $ & $7.6\pm1.0 \pm 0.6 $ & $2.1\pm0.5 \,\,^{+\,\,\,0.7}_{-\,\,\,0.2}$\\
    \hline
    \multirow{6}{*}{\Y2S} 
    & 18 -- 20        &  & $31\phantom{.} \pm4\phantom{.0} \pm 4\phantom{.0}   $ & \\
    & 20 -- 22        &  & $30\phantom{.} \pm5\phantom{.0} \pm 3\phantom{.0}   $ & \\
    & 22 -- 24        &  & $30\phantom{.} \pm6\phantom{.0} \pm 3\phantom{.0}   $ & \\
    & 24 -- 28        &  & $37\phantom{.} \pm5\phantom{.0} \,\, ^{+\,\,\,4}_{-\,\,\,8}\phantom{.0} $ & \\
    & 28 -- 40        &  & $28\phantom{.} \pm5\phantom{.0} \pm 3\phantom{.0}   $ & \\
    & 18 -- 40        &  &                                                       & $2.7\pm1.0\pm0.3$\\
    \hline
    \multirow{2}{*}{\Y3S} 
    & 24 -- 29  &  &  & $\phantom{.}34\pm8\phantom{.0}\pm 7\phantom{0.} $    \\
    & 29 -- 40  &  &  & $\phantom{.}40\pm9\phantom{.0}\,\,^{+\,\,\,5}_{-\,\,\,\,14}\phantom{.}$ \\
  \end{tabular*}
\end{table*}

In conclusion, the fractions of \ups~mesons originating from \chib~radiative decays
are measured using a~data sample collected by LHCb
at centre-of-mass energies of~7 and 8\tev, as a function of the \ups~transverse momentum
in the kinematic range $2.0<y^{\ups}<4.5$.
The~results presented in this paper supercede previous \lhcb measurements~\cite{LHCb-PAPER-2012-015}
by increasing the~statistical precision and exploiting more decay modes and higher transverse 
momentum regions.
The measurement of the 
\Y3S production fraction due to radiative \chibThreeP decays is
performed for the first time.

Assuming the mass splitting $m_{\chibtwoThreeP}-m_{\chiboneThreeP}=10.5\mevcc$,
the~mass of \chiboneThreeP~state is measured to be
\begin{equation*} 
m_{\chiboneThreeP} = 10\,511.3 \pm 1.7  \pm 2.5\mevcc,
\end{equation*}
where the first uncertainty is statistical and the second systematic.
This result is compatible and significantly more precise than 
the event yield average mass of
\chiboneThreeP~and
\chibtwoThreeP~states 
of
\mbox{$10\,530 \pm 5\pm 17\mevcc$} and 
\mbox{$10\,551 \pm 14\pm 17\mevcc$}, 
reported by the~\atlas~\cite{Aad:2011ih}
and D0~\cite{Abazov:2012gh} experiments, respectively.

\section*{Acknowledgements}

\noindent
We thank
K.-T.~Chao,
H.~Han,
V.~G.~Kartvelishvili, 
J.-P.~Lansberg
A.~K.~Likhoded,
A.~V.~Luchinsky,
S.~V.~Poslavsky  and 
H.-S.~Shao 
for inspiring and fruitful discussions
on P-wave bottomonia production.
We express our gratitude to our colleagues in the CERN
accelerator departments for the excellent performance of the LHC. We
thank the technical and administrative staff at the LHCb
institutes. We acknowledge support from CERN and from the national
agencies: CAPES, CNPq, FAPERJ and FINEP (Brazil); NSFC (China);
CNRS/IN2P3 (France); BMBF, DFG, HGF and MPG (Germany); SFI (Ireland); INFN (Italy); 
FOM and NWO (The Netherlands); MNiSW and NCN (Poland); MEN/IFA (Romania); 
MinES and FANO (Russia); MinECo (Spain); SNSF and SER (Switzerland); 
NASU (Ukraine); STFC (United Kingdom); NSF (USA).
The Tier1 computing centres are supported by IN2P3 (France), KIT and BMBF 
(Germany), INFN (Italy), NWO and SURF (The Netherlands), PIC (Spain), GridPP 
(United Kingdom).
We are indebted to the communities behind the multiple open 
source software packages on which we depend. We are also thankful for the 
computing resources and the access to software R\&D tools provided by Yandex LLC (Russia).
Individual groups or members have received support from 
EPLANET, Marie Sk\l{}odowska-Curie Actions and ERC (European Union), 
Conseil g\'{e}n\'{e}ral de Haute-Savoie, Labex ENIGMASS and OCEVU, 
R\'{e}gion Auvergne (France), RFBR (Russia), XuntaGal and GENCAT (Spain), Royal Society and Royal
Commission for the Exhibition of 1851 (United Kingdom).



\addcontentsline{toc}{section}{References}
\setboolean{inbibliography}{true}
\bibliographystyle{LHCb}
\bibliography{main,LHCb-PAPER,LHCb-CONF,LHCb-DP,LHCb-TDR,local}

\newpage

\centerline{\large\bf LHCb collaboration}
\begin{flushleft}
\small
R.~Aaij$^{41}$, 
B.~Adeva$^{37}$, 
M.~Adinolfi$^{46}$, 
A.~Affolder$^{52}$, 
Z.~Ajaltouni$^{5}$, 
S.~Akar$^{6}$, 
J.~Albrecht$^{9}$, 
F.~Alessio$^{38}$, 
M.~Alexander$^{51}$, 
S.~Ali$^{41}$
% ...

{\footnotesize \it
% ...
$ ^{5}$Clermont Universit\'{e}, Universit\'{e} Blaise Pascal, CNRS/IN2P3, LPC, Clermont-Ferrand, France\\
$ ^{6}$CPPM, Aix-Marseille Universit\'{e}, CNRS/IN2P3, Marseille, France\\
% ...
$ ^{9}$Fakult\"{a}t Physik, Technische Universit\"{a}t Dortmund, Dortmund, Germany\\
% ...
$ ^{37}$Universidad de Santiago de Compostela, Santiago de Compostela, Spain\\
$ ^{38}$European Organization for Nuclear Research (CERN), Geneva, Switzerland\\
% ...
$ ^{41}$Nikhef National Institute for Subatomic Physics, Amsterdam, The Netherlands\\
% ...
$ ^{46}$H.H. Wills Physics Laboratory, University of Bristol, Bristol, United Kingdom\\
% ...
$ ^{51}$School of Physics and Astronomy, University of Glasgow, Glasgow, United Kingdom\\
$ ^{52}$Oliver Lodge Laboratory, University of Liverpool, Liverpool, United Kingdom\\
% ...
}
\end{flushleft}

%
%

\end{document}